\documentclass[%
 reprint,
superscriptaddress,
 amsmath,amssymb,
 aps,
 prb,
]{revtex4-2}

\usepackage{graphicx}
\usepackage{dcolumn}
\usepackage{bm}
\usepackage{hyperref}
\usepackage[shortcuts]{extdash}

\usepackage{xcolor}
\usepackage{textgreek}

\begin{document}

\title{Magnetic field-induced weak-to-strong-link transformation in patterned superconducting films}

\author{Davi A. D. Chaves}
    \email{davi@df.ufscar.br}
\affiliation{Departamento de Física, Universidade Federal de São Carlos, 13565-905 São Carlos, SP, Brazil}

\author{M. I. Valerio-Cuadros}
\altaffiliation[Current address: ]{Departamento de Física, Universidade Estadual de Maringá, 87020-900 Maringá, PR, Brazil}
\affiliation{Departamento de Física, Universidade Federal de São Carlos, 13565-905 São Carlos, SP, Brazil}

\author{L. Jiang}
\affiliation{School of Aeronautics, Northwestern Polytechnical University, Xi'an 710072, China}
\affiliation{Experimental Physics of Nanostructured Materials, Q-MAT, CESAM, Université de Liège, B-4000 Sart Tilman, Belgium}

\author{E. A. Abbey}
\affiliation{Departamento de Física, Universidade Federal de São Carlos, 13565-905 São Carlos, SP, Brazil}

\author{F. Colauto}
\affiliation{Departamento de Física, Universidade Federal de São Carlos, 13565-905 São Carlos, SP, Brazil}

\author{A. A. M. Oliveira}
\affiliation{Instituto Federal de Educa\c{c}\~ao, Ci\^encia e Tecnologia de S\~ao Paulo, Campus S\~ao Carlos, 13565-905, S\~ao Carlos, SP, Brazil}

\author{A. M. H. Andrade}
\affiliation{Instituto de F\'isica, Universidade Federal do Rio Grande do Sul, 91501-970 Porto Alegre, RS, Brazil.}

\author{L. B. L. G. Pinheiro}
\affiliation{Departamento de Física, Universidade Federal de São Carlos, 13565-905 São Carlos, SP, Brazil}
\affiliation{Instituto Federal de Educa\c{c}\~ao, Ci\^encia e Tecnologia de S\~ao Paulo, Campus S\~ao Carlos, 13565-905, S\~ao Carlos, SP, Brazil}

\author{T. H. Johansen}
\affiliation{Department of Physics, University of Oslo, P.O. Box 1048 Blindern, 0316 Oslo, Norway}

\author{C. Xue}
\affiliation{School of Mechanics, Civil Engineering and Architecture, Northwestern Polytechnical University, Xi'an 710072, China}

\author{Y.-H. Zhou}
\affiliation{Department of Mechanics and Engineering Sciences, Lanzhou University, Key Laboratory of Mechanics on Disaster and Environment in Western China, Ministry of Education of China, Lanzhou 730000, China}
\affiliation{School of Aeronautics, Northwestern Polytechnical University, Xi'an 710072, China}

\author{A. V. Silhanek}
\affiliation{Experimental Physics of Nanostructured Materials, Q-MAT, CESAM, Université de Liège, B-4000 Sart Tilman, Belgium}

\author{W. A. Ortiz}
\affiliation{Departamento de Física, Universidade Federal de São Carlos, 13565-905 São Carlos, SP, Brazil}

\author{M. Motta}
\email{m.motta@df.ufscar.br}
\affiliation{Departamento de Física, Universidade Federal de São Carlos, 13565-905 São Carlos, SP, Brazil}

\date{\today}

\begin{abstract}
     Ubiquitous in most superconducting materials and a common result of nanofabrication processes, weak-links are known for their limiting effects on the transport of electric currents. Still, they are at the root of key features of superconducting technology. By performing quantitative magneto-optical imaging experiments and thermomagnetic model simulations, we correlate the existence of local maxima in the magnetization loops of FIB-patterned Nb films to a magnetic field-induced weak-to-strong-link transformation increasing their critical current. This phenomenon arises from the nanoscale interaction between quantized magnetic flux lines and FIB-induced modifications of the device microstructure. Under an ac drive field, this leads to a rectified vortex motion along the weak-link. The reported tunable effect can be exploited in the development of new superconducting electronic devices, such as flux pumps and valves, to attenuate or amplify the supercurrent through a circuit element, and as a strategy to enhance the critical current in weak-link-bearing devices.
\end{abstract}

\maketitle


\section{Introduction}

Nanoscale patterning of superconducting films enables the optimization and control of distinct material properties~\cite{harada1996direct,silhanek2003guided,dobrovolskiy2012electrical,shaw2012properties,wang2013enhancing,dobrovolskiy2017abrikosov,kalcheim2017dynamic,dobrovolskiy2017pinning,colauto2021controlling}. Moreover, patterning allows exploring different phenomena rising from the rich physics of superconducting weak-links (WLs)~\cite{likharev1979superconducting,moshchalkov2010nanoscience,fornieri2019evidence,holzman2019superconducting}. Today, patterned superconducting structures find applications in memories~\cite{murphy2017nanoscale,chen2020miniaturization,ilin2021supercurrent,chaves2023nanobridge}, diodes~\cite{lyu2021superconducting,golod2022demonstration}, quantum batteries~\cite{strambini2020josephson}, and phase shifters~\cite{yamashita2020pi,golod2021reconFigurable}, serving as the backbone of superconducting electronics.

Meanwhile, focused-ion beam (FIB) milling is a prominent fabrication technique with nanometric spatial resolution~\cite{volkert2007focused,wyss2022magnetic,sigloch2022direct}. However, FIB unavoidably introduces defects along the patterned regions, arising from ionic implantation~\cite{datesman2005gallium,deleo2016thickness,singh2018influence}, locally increasing vortex pinning in superconductors~\cite{dobrovolskiy2012electrical,dobrovolskiy2017pinning,aichner2019ultradense}. Recently, we investigated the effects of a single FIB-milled WL on the properties of prototypical low-critical temperature Nb films, revealing the existence of local maxima in their magnetization loops~\cite{valerio2021superconducting}.

Weak-links are also an important challenge in the optimization of large-scale applications of high-temperature superconductors (HTSC)~\cite{larbalestier2001high,foltyn2007materials}. Particularly, the overall critical current density ($J_c$) of HTSC is affected by an angular-dependent deterioration of $J_c$ along grain boundaries~\cite{hilgenkamp2002grain,graser2010grain}, motivating the development of specialized fabrication techniques~\cite{reade1992laser,ijima1993structural,wu1995properties,goyal1996high}. Improving critical currents in HTSC and low-critical-temperature materials remains an active research topic~\cite{diez2020high,ahmad2021enhanced,rocci2020large,caffer2021optimum,algarni2021enhanced,chaves2021enhancing}.

In this study, we investigate magnetic flux penetration and shielding currents in FIB-patterned Nb films containing either a single or four artificial WLs. Our results explain the origin of local maxima in the magnetization loops, revealing a field-induced transformation from a weak-link to a strong-link behavior by the enhancement of $J_c$ across the WL.

\begin{figure*}[t!]
    \includegraphics[width=\linewidth]{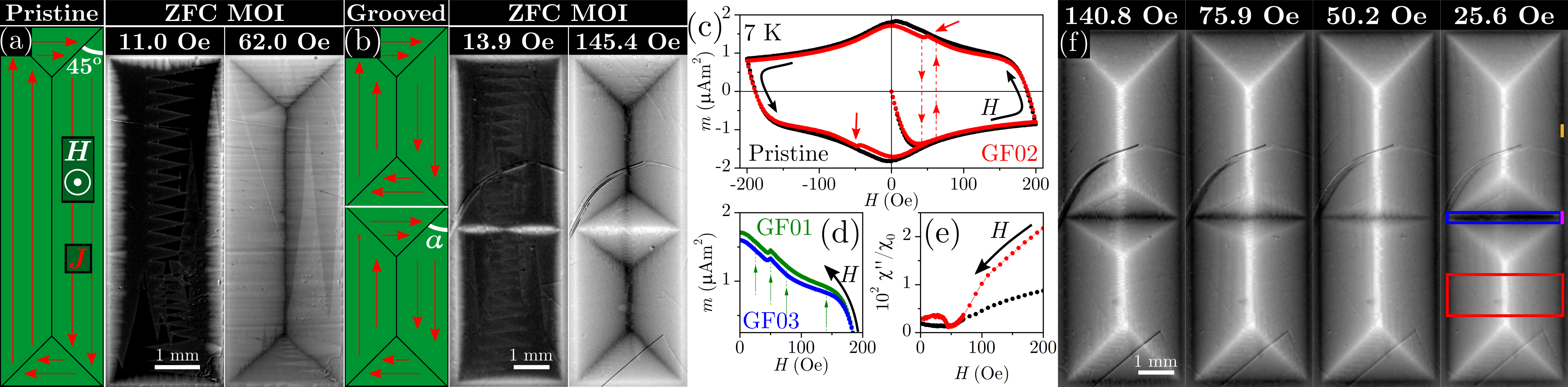}
    \caption{Representation and MOI at 7~K of fully flux-penetrated (a) pristine and (b) grooved GF01 films under an increasing $H$ after ZFC. Red arrows represent the direction of $\bm{J}$, $\alpha$ is the angle between the d-line and the edge. (c) $m(H)$ at 7~K for the pristine and GF02 samples. (d) The positive field-decreasing branch of $m(H)$ for GF01 and GF03. (e) $\chi''$ at 7~K as $H$ is reduced for the pristine and GF02 samples. The probe field has an amplitude of 3~Oe and a frequency of 100~Hz. The data is normalized by the 2~K, zero-field in-phase response for each sample, $\chi_0$. (f) MOI of GF01 at 7~K for the decreasing values of $H$ indicated by vertical arrows in the main panel in (d). Colored rectangles represent areas used in the investigation in Fig.~\ref{Fig:dLineMovement}(b-c).}
  \label{Fig:FluxPenetration-MagPeaks}
\end{figure*}

\section{Methods}

Three 180-nm-thick Nb films were patterned with a single shallow groove using FIB doses of 0.1, 0.2, and 0.3~nC/\textmu m$^2$ (GF01, GF02, and GF03, respectively). The samples have areas of 3~mm$^2$ (3$\times$1) in the case of GF01 and 2~mm$^2$ (2.5$\times$0.8) for the other films. This process creates WLs comprised of thinner Nb regions pervaded by defects arising from Ga$^+$ implantation, as shown in Appendix~\ref{Sc:AppA}. The groove depths range from 4.2~nm to 14.6~nm. Additionally, a 2.5~$\times$~2.5~mm$^2$ 300-nm-thick Nb film labeled GF01+ was deposited and patterned with two perpendicular grooves using a FIB dose of 0.1~nC/\textmu m$^2$ thus exhibiting four distinct WLs in a cross pattern. A pristine film with critical temperature ($T_c$) of 9.0~K is also studied. Sample preparation details are given in Ref.~\citenum{valerio2021superconducting}. Global dc magnetic responses are obtained in a QD MPMS-5S in the perpendicular geometry. Quantitative magneto-optical imaging (MOI) is used to reveal the local out-of-plane magnetic flux density ($B$) and current density ($\bm{J}$) in the films~\cite{shaw2018quantitative,meltzer2017direct}.

Furthermore, we employed the thermomagnetic model (TM) to simulate the macroscopic magnetic flux density distribution in the patterned films~\cite{vestgarden2011dynamics}. For that, the heat diffusion equation
\begin{equation} \label{Eq:HeatDiffusion}
dc\dot{T}=d\kappa\nabla^2 T-h(T-T_{0})+\bm{j}\cdot\bm{E}
\end{equation}
is solved considering Maxwell's equations and that the superconductor material properties are given by
\begin{equation} \label{E1}
\bm{E}=\left\{
\begin{array}{ll}
\rho_{n}(j/j_{c})^{n-1}\bm{j}/d &\text{if}~j<j_{c}~\text{and}~T<T_{c},\\
\rho_{n}\bm{j}/d &\text{otherwise},
\end{array}
\right.
\end{equation}
where $\bm{E}$ is the electric field, $\kappa$ the thermal conductivity, $c$ the specific heat, $h$ the coefficient for heat removal to the substrate, $\rho_{n}$ the normal state resistivity, $d$ is the film thickness, $\bm{j}$ the sheet current, $j_c$ the critical sheet current, and $n$ the flux creep exponent~\cite{vestgarden2011dynamics}. The simulation parameters used in this work follow Ref.~\cite{jiang2020selective}.

\section{Results \& discussion}

The left panel of Fig.~\ref{Fig:FluxPenetration-MagPeaks}(a) schematically represents a superconducting film fully penetrated by flux under a perpendicular magnetic field $H$ after zero-field cooling (ZFC)~\cite{schuster1994observation,schuster1995discontinuity}. We observe five dark discontinuity lines, or d-lines, shielding flux more efficiently where $\bm{J}$ changes its direction. The magneto-optical (MO) image at 11.0~Oe demonstrates that flux penetrates from the edges toward the center of the pristine film as ${H}$ is increased. This is manifested by bright, flux-filled regions surrounding the dark, shielded central portion of the sample. At 62.0~Oe, we observe the expected domains for the fully penetrated film. Although the resolution of MOI reveals an apparently continuous flux front, the superconductor is in fact permeated by quantized flux lines, or vortices~\cite{Mints1981,Blatter1994}.

In contrast, a groove across the shortest symmetry line of the film creates a WL, defining two apparently disconnected pristine regions (PR). Since the WL has a lower $J_c$ than the PR, $\bm{J}$ needs to bend away from the groove, resulting in the central diamond-shaped domain represented in Fig.~\ref{Fig:FluxPenetration-MagPeaks}(b)~\cite{polyanskii1996magneto,johansen2019transparency,colauto2021measurement}. In the limiting case when the depicted angle $\alpha$~=~45$^\circ$, no current is able to flow through the WL, i.e., $J_c^{\text{WL}} = 0$. MOI for GF01 at 13.9~Oe reveals that the WL is fully penetrated\textemdash a consequence of its weaker shielding capacity. At 145.4~Oe, the diamond-shaped domain can be clearly distinguished. Contrary to the representation, MOI shows $\alpha>45^\circ$ and a faded dark vertical d-line appears inside the diamond. These effects happen because $J_c^{\text{WL}} > 0$, meaning that a fraction of $\bm{J}$ is able to flow through the WL. The dark scratches above the diamond-shaped and bottom triangular domains are defects on the MO indicator and do not interfere with the flux penetration into the sample.

\begin{figure}[h]
    \includegraphics[width=\linewidth]{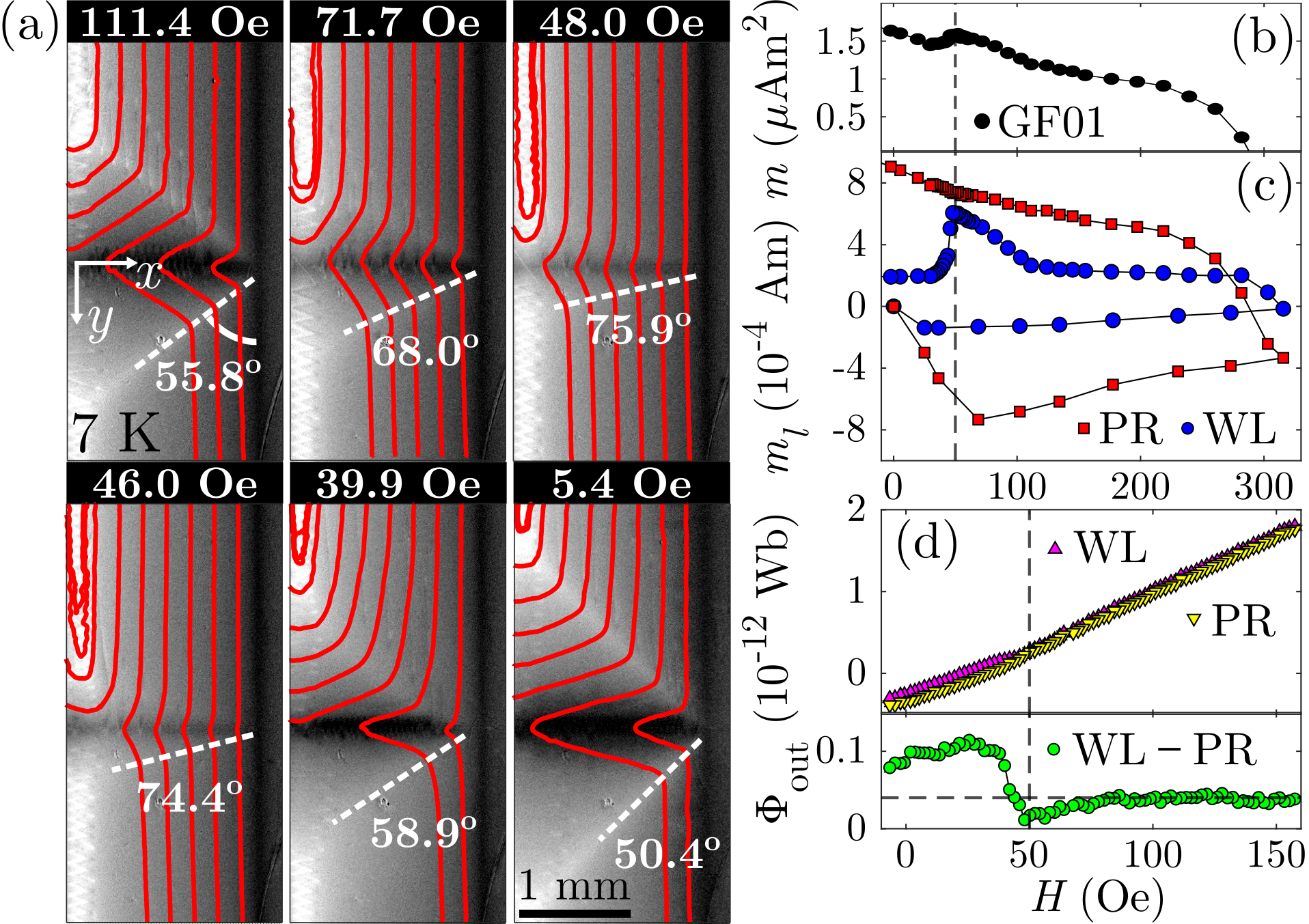}
    \caption{(a) MOI of GF01 at 7~K as $H$ is reduced from full penetration. The $\bm{J}$ distribution is shown as red streamlines. (b) $m(H)$ for GF01 evaluated by MOI. (c) $m_l(H)$ averaged over different regions of the sample. (d) Upper panel shows $\Phi_{\text{out}}$ besides the WL and the PR. Lower panel highlights the difference in $\Phi_{\text{out}}$ between the two regions. The quantities are obtained for the color-matched regions shown in Fig.~\ref{Fig:FluxPenetration-MagPeaks}(d).}
  \label{Fig:dLineMovement}
\end{figure}

\begin{figure*}[t]
    \includegraphics[width=\linewidth]{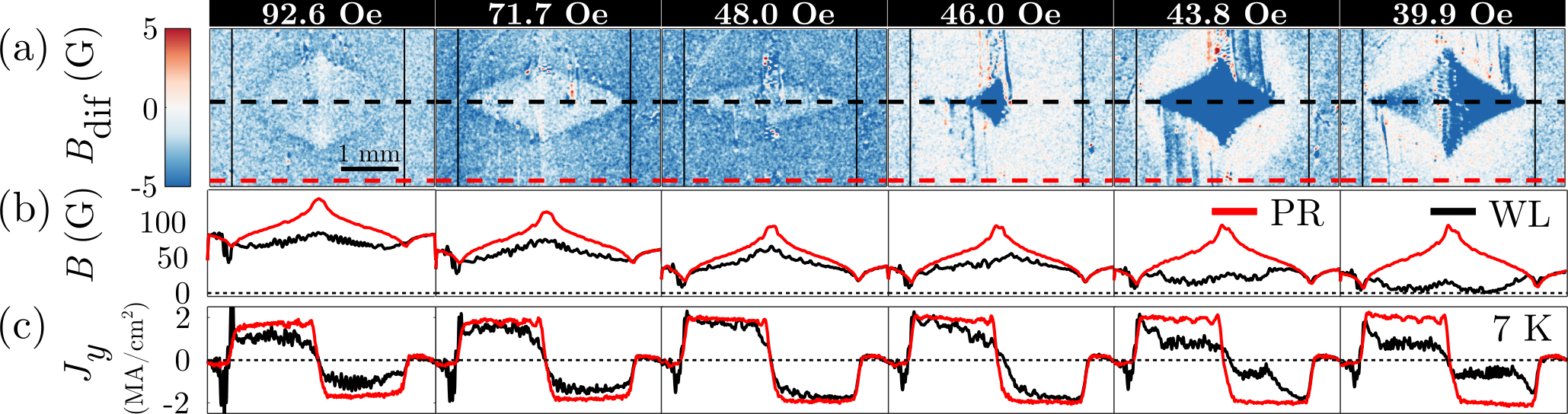}
    \caption{(a) $B_{\text{dif}}$ distribution around the region of d-line movement for GF01 at 7~K at different $H$. Thin vertical lines represent the edges of the film. (b) $B$ and (c) $J_y$ profiles along the WL and PR, evaluated respectively at the black and red dashed lines in panel (a).}
  \label{Fig:Profiles}
\end{figure*}

Figure~\ref{Fig:FluxPenetration-MagPeaks}(c) shows complete magnetic moment hysteresis loops, $m(H)$, at 7~K for the pristine film and GF02. The pristine behavior matches that expected for a type-II superconductor presenting a flux-dependent critical current density, $J_c(B)$~\cite{kim1963magnetization,shantsev2000thin}. In contrast, GF02 exhibits local maxima in the positive and negative decreasing-field branches of $m(H)$. Figure~\ref{Fig:FluxPenetration-MagPeaks}(d) reveals the same feature around 50~Oe for GF01 and GF03. Out-of-phase ac magnetic susceptibility ($\chi''$) measurements in Fig.~~\ref{Fig:FluxPenetration-MagPeaks}(e) demonstrate that this effect is sensible to ac excitations. For sufficiently large probe fields, a dip in $\chi''$ occurs around the same $H$ values at which $m(H)$ peaks for GF02. This ac experiment is equivalent to performing a minor $m(H)$ loop, as indicated by dashed arrows in Fig.~\ref{Fig:FluxPenetration-MagPeaks}(c). Thus, at the upper branch of the loop, while the probe field decreases the total applied field, the sample experiences the same phenomenon giving rise to the $m(H)$ peak.

We elect GF01 to illustrate our investigation of what leads to this phenomenon. GF01 contains a 4.2-nm-deep groove with an 800~nm width\textemdash see Appendix~\ref{Sc:AppA}. Figure~\ref{Fig:FluxPenetration-MagPeaks}(f) shows a series of MO images as ${H}$ is decreased from full penetration. At 140.8~Oe, flux penetration is similar to that of Fig~\ref{Fig:FluxPenetration-MagPeaks}(b). The WL now appears in dark contrast, because $H$ was reduced from 315.6~Oe, and the WL shows a lower, but positive flux density due to its weaker shielding capability. However, as ${H}$ approaches 50~Oe, the d-lines forming the central diamond-shaped domain move toward the groove\textemdash as if they were closing. The image at 50.2~Oe resembles a pristine film, as the diamond shape practically vanishes. If $H$ is further reduced, the d-lines move away from the groove, as if they were reopening, reestablishing the diamond-shaped domain, as seen at 25.6~Oe. This analysis demonstrates that MOI is an ideal tool for this study, revealing details of the local flux penetration and current distribution that are hidden in global magnetometry.

The observed d-line movement is associated with an enhancement of the transport properties in the WL. This is demonstrated in Fig.~\ref{Fig:dLineMovement}(a). As the d-lines close, $\alpha$ increases, and $\bm{J}$ changes accordingly\textemdash streamlines that initially did not cross the WL become straighter and have a higher density across the WL. This trend is maximized at 48.0~Oe, when the diamond-shaped domain almost fades out. Thus, after the sample is fully penetrated and $H$ is reduced to a specific value, the WL behaves as a strong-link, allowing currents to flow through it largely unaffected. At 46.0~Oe and as $H$ is further reduced, the d-lines reappear, reestablishing the WL behavior. As $J_c$ is proportional to the height of the $m(H)$ curve~\cite{johansen1995critical}, this notion explains the observed local peak as an increase in the overall $J_c$ due to an increase in $J_c^{\text{WL}}$.

In Fig.~\ref{Fig:dLineMovement}(b), we evaluate the magnetic moment of GF01 via MOI as $m = \sum_y m_l(y)l_{px} = l_{px}\sum_y 2\int_0^{w/2} xj_y(x)dx$, where $m_l(y)$ is $m$ per unit length at a position $y$, $l_{px}$ is the image pixel size, $w$ is the film width, $j_y = J_yd$ is the sheet current flowing perpendicular to the WL, and $d$ is the film thickness~\cite{brandt1993Type,shantsev1999central}. The results reproduce those in Fig.~\ref{Fig:FluxPenetration-MagPeaks}(c). Figure~\ref{Fig:dLineMovement} also presents quantitative results obtained along the four color-matched regions in Fig.~\ref{Fig:FluxPenetration-MagPeaks}(d). In Fig.~\ref{Fig:dLineMovement}(c), we average $m_l$ over the WL (blue) and on a selected portion of the PR (red), confirming that the peak in $m(H)$ manifests an effect tied to the WL. Moreover, we gauge the flux behavior just outside the sample, $\Phi_{\text{out}}$. Figure~\ref{Fig:dLineMovement}(d) shows that, as $H$ is reduced, $\Phi_{\text{out}}$ is everywhere higher just beside the WL (pink) than beside the PR (yellow) for equal areas. However, as the d-lines are closing, $\Phi_{\text{out}}$ decreases close to the WL while it is unaffected beside the PR, lessening the difference between $\Phi_{\text{out}}$ along the two regions (green). This indicates that fewer vortices are able to escape the sample through the WL at these fields. When the d-lines reopen below 50~Oe, $\Phi_{\text{out}}$ is enhanced besides the groove, indicating a large amount of flux is expelled through the edges of the WL. In the ac experiment, if the device is properly field-biased, this implies a rectified vortex motion characterized by the directed flux movement along the WL albeit vortices are subjected to a zero-mean excitation~\cite{Wambaugh1999,Olson2001,Togawa2005,Clecio2006a,Clecio2006b}.

Figure~\ref{Fig:Profiles} locally resolves the behavior of $B$ and $J$ for GF01. Figure~\ref{Fig:Profiles}(a) depicts the differential flux density, $B_{\text{dif}}$, for GF01. It captures the $B$ variation due to the variation of $H$ (kept around $-2$~Oe) by subtracting the target MO image $n$ by the one taken at the previous field step, i.e. $B_{\text{dif}}~=~B_n-B_{n-1}$. In the first three $B_{\text{dif}}$ images, we visualize the closing of the d-lines by the shrinking of a brighter diamond-shaped inner region. This indicates that $B$ decreases less in that region than outside the d-lines. Therefore, flux is pushed toward the WL as the d-lines close. This agrees with the fact that flux is unable to cross d-lines~\cite{Schuster1994}. At 46.0~Oe, a dark-colored region confirms flux is intensely expelled from the WL. At 43.8~Oe, the bright halo centered in the groove reveals that flux is still being pushed away from the inner part of the sample. For 39.9~Oe and further, the flux pushed out of the WL is systematically reduced.

Figure~\ref{Fig:Profiles}(b) shows the evolution of $B$ along the WL (in black) and the PR (red). From 92.6 to 48.0~Oe, the black profile decreases less and becomes increasingly similar to the red one, indicating that the flux pushed inward by the d-line movement is partially retained by the WL. This is related to an increased vortex pinning potential in the WL due to its reduced thickness and FIB-induced defects, which act as pinning centers and locally suppress $T_c$~\cite{dobrovolskiy2012electrical,valerio2021superconducting}\textemdash see Appendix~\ref{Sc:AppA}. The observed relative increase in $B$ corroborates the increase in $m_l$ across the WL seen in Fig.~\ref{Fig:dLineMovement}(c). Starting at 46.0~Oe, trapped flux is vigorously pushed away from the WL and the $B$ profiles become distinct again. As $B$ is still positive at 46.0~Oe, the maximum in $m(H)$ is not related to a net neutral flux in the WL. Here, we argue that the increased vortex pinning in the WL reduces energy dissipation coming from flux line movement, thus increasing $J_c^{\text{WL}}$ within that specific $H$ range. A similar behavior exists in HTSC~\cite{evetts1988relation,dimos1990superconducting,shantsev1999central,hogg2001angular,thompson2004self,palau2004simultaneous,palau2007simultaneous}. The fact that the associated d-line movement was not previously observed in YBCO bicrystalline films~\cite{polyanskii1996magneto,guth2004inhomogeneous,born2004self,palau2007simultaneous} may be related to the absence of the FIB-added pinning centers and the relatively higher $H$ for which the peak occurs in HTSC, rendering $J_c(B)$-dependent effects less prominent~\cite{shantsev2000thin}.

Figure~\ref{Fig:Profiles}(c) presents an analysis of $J$ perpendicular to the WL, $J_y$. Initially, there is a gradual increase in the current able to cross the WL. At 48.0~Oe, $J_y$ across the WL closely matches that flowing in the PR\textemdash as GF01 behaves almost as a pristine film. This trend is reversed when the d-lines reappear and a sharp decrease in $J_y$ is observed beginning in the center of the WL. The increased current density, paired with the higher vortex concentration at the center of the WL results in a repulsive force that eventually overcomes the pinning force, thus relieving the magnetic pressure built up in the WL and inducing a strong-to-weak-link transformation. Hence, $J_c$ for a field-biased grooved sample subjected to ac fields alternates between maxima and minima in intervals dictated by the drive field. Appendix~\ref{Sc:AppB} depicts the behavior of $B$ and $J_y$ at several applied fields, corroborating their relationship to the d-line movement.

With the experimental input, we now turn to numerical simulations based on the thermomagnetic model, which allow for the comprehension of subtle distinctions in specimen behavior arising from a $B$-dependent $J_c$~\cite{jing2016influences,jing2017numerical,jiang2020selective}. The simulated Nb sample shares the geometry of GF01, except for the groove width, set to 45~\textmu m to improve computational performance. Notably, the higher density of pinning sites introduced by Ga implantation in the WL is not considered. A reduction of $\sim$15\% in the zero-field $J_c^{\text{WL}}$ is induced by reducing $T_c$ in the WL, as experimentally observed\textemdash see Appendix~\ref{Sc:AppA} and Ref.~\citenum{valerio2021superconducting}.

\begin{figure}[h]
    \includegraphics[width=\linewidth]{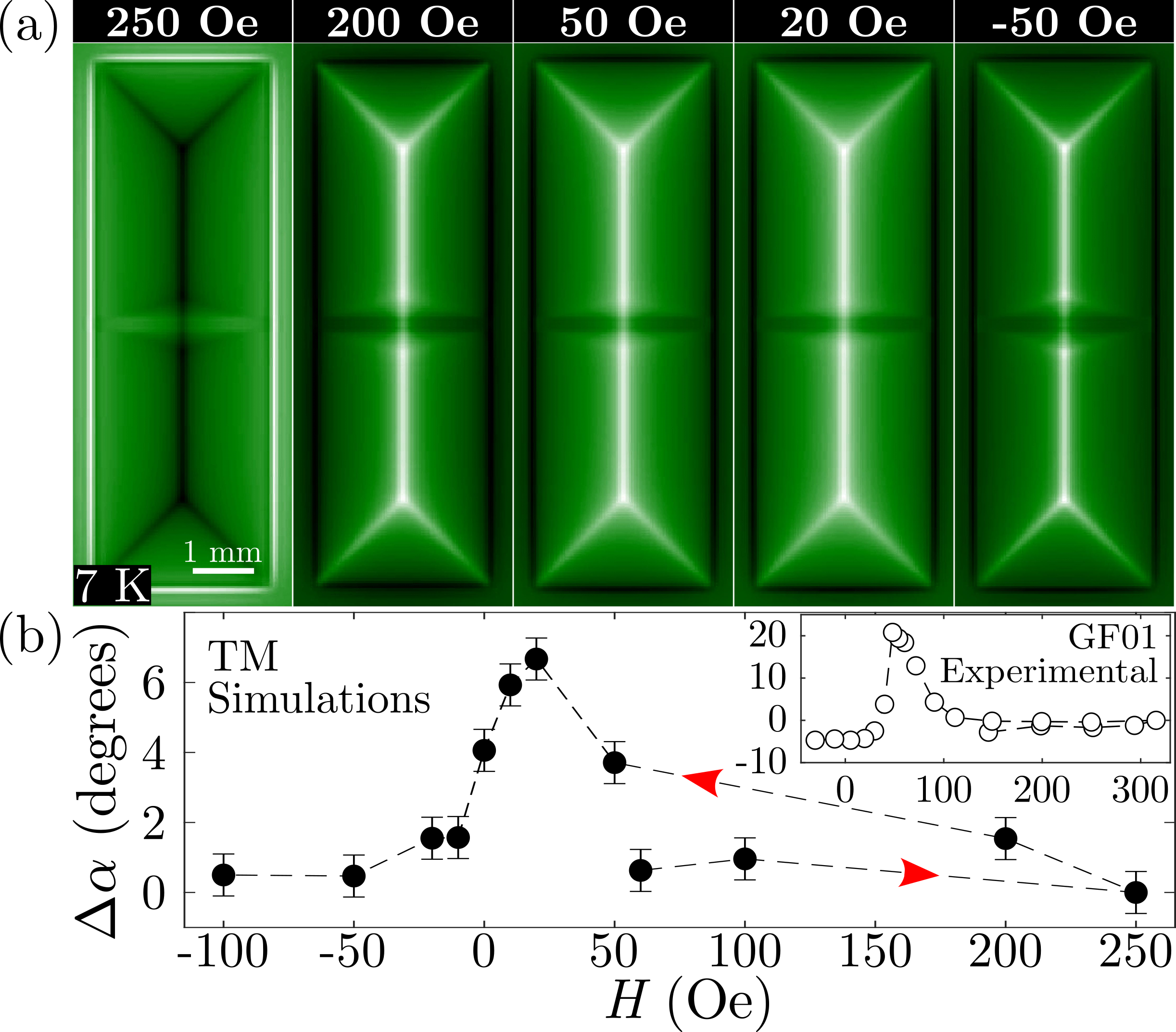}
    \caption{(a) $B$ distribution {at 7~K} captured by TM simulations as $H$ is reduced from 250~Oe. (b) $\Delta\alpha(H)$ for the simulated sample. Inset: $\Delta\alpha(H)$ for GF01.}
  \label{Fig:Simulation}
\end{figure}

\begin{figure}[h]
    \includegraphics[width=\linewidth]{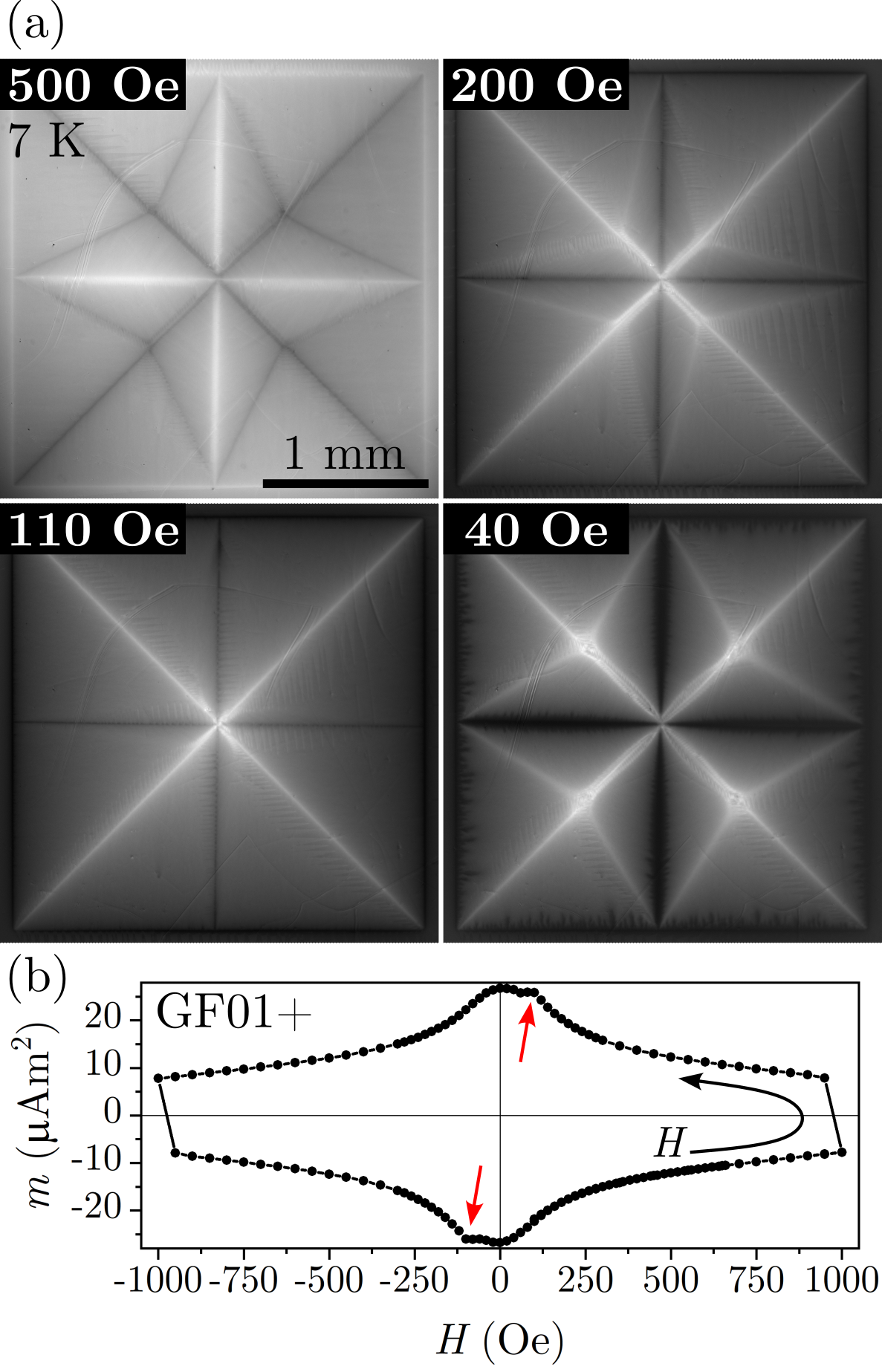}
    \caption{(a) MOI and (b) $m(H)$ of GF01+ at 7~K. The MO images are acquired as $H$ is reduced from 500~Oe, at which the film is fully flux penetrated. Red arrows indicate the local maxima in $m(H)$.}
  \label{Fig:Cross}
\end{figure}

Figure~\ref{Fig:Simulation} describes the TM results obtained when considering $j_c=j_c(T,B)=j_{c0}(1-T/T_c)(B_0/(B+B_0))$, with $B_0=\mu_0j_{c0}/\pi$~\cite{kim1962critical}. Disorder is introduced by lowering $j_{c0}$ for randomly selected grid points~\cite{vestgarden2011dynamics}. First, $H$ is increased up to 250~Oe after ZFC to 7~K. The resulting flux distribution in Fig.~\ref{Fig:Simulation}(a) qualitatively matches that of GF01 in the full penetration state. Then, $H$ is progressively reduced. At 50~Oe, we notice the shrinking of the diamond-shaped domain, which is maximum at 20~Oe. This is quantitatively corroborated by Fig.~\ref{Fig:Simulation}(b), which depicts $\Delta\alpha(H) = \alpha(H=250~\text{Oe}) - \alpha(H)$. The inset of Fig.~\ref{Fig:Simulation}(b) confirms that the simulations reproduce the d-line movement observed experimentally. The lesser closure of the d-lines is likely due to the larger width of the simulated groove. Finally, the d-lines reappear, as demonstrated in Fig.~\ref{Fig:Simulation}(a) at $-50$~Oe. TM simulations at 7 K indicate a lower threshold of $J_{c0}^{\text{WL}}/J_{c0} \sim 64\%$ for the occurrence of the reported phenomenon. This is higher than the $\sim40\%$ ratio for GF01 at 7 K due to the different geometry and the irradiation-induced pinning centers. A lower bound for $J_{c0}^{\text{WL}}/J_{c0}$ corroborates experimental observations\textemdash see Appendix~\ref{Sc:AppA}. Additional simulations, conducted for $J_c$ independent of $B$, did not reproduce the d-line movement, reinforcing the $J_c(B)$-dependency role in the observed phenomenon.

Finally, Fig.~\ref{Fig:Cross}(a) presents MO results for GF01+. The weak-links appear in bright contrast at the image captured at the maximum applied field of 500~Oe. Enveloping the WLs, a set of four quadrilateral domains define a remarkable four-pointed star-like pattern~\cite{colauto2021measurement}. As the field is reduced to 200~Oe, such a pattern systematically fades out, as the d-lines close around the WLs\textemdash see Appendix~\ref{Sc:AppB}. At 110~Oe, the observed flux distribution matches that of a pristine square film~\cite{johansen1995critical}, as all four WLs simultaneously behave as strong-links. This coincides with a local peak in $m(H)$, as revealed by Fig.~\ref{Fig:Cross}(b). Further $H$ reduction will cause the d-lines to reopen, reestablishing the weak-link behavior and the star-like pattern. Therefore, GF01+ behaves completely analogously to GF01 and provides further evidence of the phenomenon reproducibility while revealing it can be scaled up for samples containing multiple WLs.

\section{Conclusions}

In summary, combining MOI and TM simulations, we demonstrate that the presence of local maxima in the magnetization loops of FIB-patterned Nb films is associated with a field-induced transformation of the sample behavior from weak-to-strong-link. This effect can be tuned by temperature~\cite{valerio2021superconducting} and the amount of defects introduced along the WL. As the transformation is associated with nanoscale interactions between vortices and defects, the phenomenon should also be observed in micro and nanometric devices, more typical size scales for superconducting electronics. Possible applications can take advantage of the vortex rectification and tunable flux escape along the WL in flux pumps or valves. The overall $J_c$ modulation can be explored for constant or periodic signal attenuation or amplification in selected circuit components. Additionally, the targeted inclusion of pinning centers along grain boundaries of weak-link-bearing materials may also lead to the improvement of operating conditions for different superconducting technologies, resulting in effectively higher critical currents for HTSC devices.

\acknowledgments{This work was partially supported by Coordenação de Aperfeiçoamento de Pessoal de Nível Superior -- Brasil (CAPES) -- Finance Code 001, the São Paulo Research Foundation (FAPESP, Grant No. 2021/08781-8), and the National Council for Scientific and Technological Development (CNPq, Grants No. 431974/2018-7, No. 316602/2021-3, and No. 309928/2018-4). C. X. acknowledges the support by the National Natural Science Foundation of China (Grants No. 11972298). L. J. was supported by the China Scholarship Council. The authors thank the Laboratory of Structural Characterization (LCE/DEMa/UFSCar) for the EDS and AFM measurements. Research supported by LNNano -- Brazilian Nanotechnology National Laboratory (CNPEM/MCTI) during the use of the Device Manufacturing open access facility.}

\appendix

\section{Morphology, structure, and superconducting properties}
\label{Sc:AppA}

\begin{figure*}
    \includegraphics[width=\linewidth]{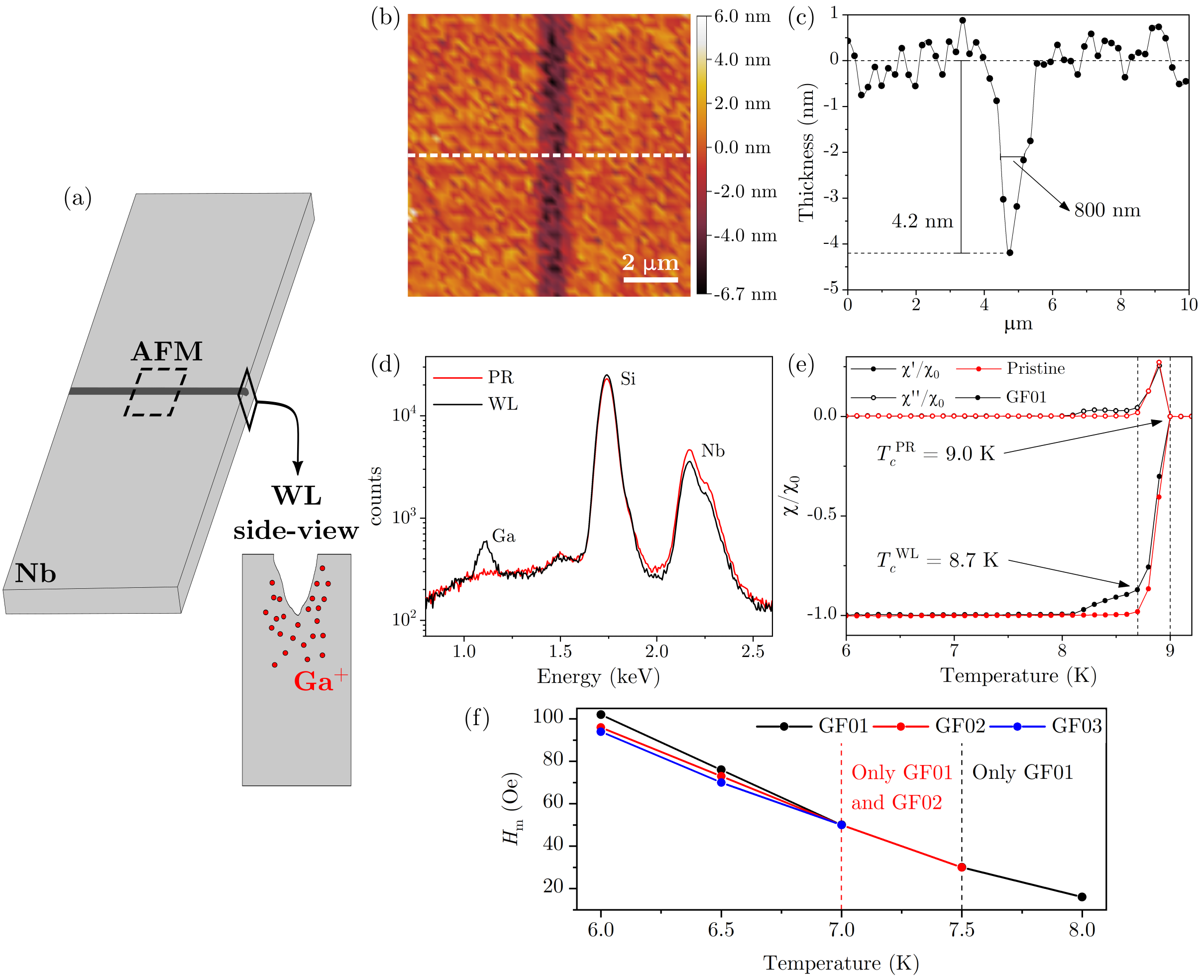}
    \caption{(a) Schematic representation of the grooved Nb film. In detail, the lateral cross-section of the sample highlights the WL profile. Red circles represent the damage caused by Ga implantation. (b) AFM of a 10$\times$10~\textmu m$^2$ area around the central WL of GF01, as represented in (a). The values are shown in comparison to the film's mean thickness to clearly show the groove depth. (c) Thickness profile centered along the dashed white line in (b).  (d) EDS spectra of GF02 measured in the PR (in red) and the WL (in black). The results reveal the existence of a Ga {K\textalpha} emission peak only in the WL. (e) Temperature-dependent AC-susceptibility for GF01 and the pristine sample, highlighting the different superconducting transition temperatures for the PR and the WL. (f) The variation of $H_m$ as a function of temperature for GF01, GF02, and GF03. From left to right, the dashed lines delimited temperatures for which the peak is only observed for GF01 and GF02 and only for GF01.}
  \label{Fig:Suppl}
\end{figure*}

The Nb samples are deposited by magnetron sputtering on a SiO$_2$ substrate, as described in Ref.~\cite{valerio2021superconducting}. The single-WL samples are schematically represented in Fig.~\ref{Fig:Suppl}(a), where the grooved region appears in dark in the center of the film, spanning the whole sample's width. The lower right corner shows a sketch of the weak-link lateral profile, representing the FIB-milled-induced defects arising from Ga implantation as red circles. SRIM simulations~\cite{SRIM} indicate Ga ions penetrate less than 30~nm into the Nb film. As described in Refs.~\cite{valerio2021superconducting,datesman2005gallium,deleo2016thickness}, the Ga implantation is an unavoidable consequence of FIB-milling, leading to a suppression of the Nb sample superconducting properties.

Figure~\ref{Fig:Suppl}(b) presents Atomic-Force Microscopy (AFM) results for GF01 centered around the WL, confirming the presence of the groove. The AFM measurements were conducted on a Digital Instruments Nanoscope V. The thickness values are presented with respect to the average height of the Nb film, which highlights the groove depth. In Fig.~\ref{Fig:Suppl}(c), a thickness profile is plotted centered along the white dashed line in panel (b). The points are averaged for a 1$\times$10~\textmu m$^2$ area. The profile reveals a maximum groove depth of 4.2~nm for GF01, as well as an 800~nm width at half depth.

Figure~\ref{Fig:Suppl}(d) confirms the Ga implantation inside the WL. This is accomplished by performing an Energy Dispersive X-ray Spectrometry (EDS), conducted using a Philips XL-30 FEG Scanning Electron Microscope. We take EDS spectra of GF02 of both the pristine region (PR) and the WL, which are shown in red and black, respectively. For the PR, EDS reveals characteristic {K\textalpha} emission peaks for Si (due to the substrate) and Nb. However, when the WL is gauged, an additional emission line due to the presence of Ga is observed.

Figure~\ref{Fig:Suppl}(e) demonstrates the deterioration of the superconducting properties in the WL due to the milling process. It shows the temperature-dependency of both in-phase ($\chi'$) and out-of-phase ($\chi''$) AC-susceptibility components for GF01, in black, and the pristine sample, in red. The results were obtained on a SQUID-based Quantum Design MPMS-5S magnetometer and the curves for each sample are normalized by their respective $\chi'$ Meissner state plateau, $\chi_0$. A 100~Hz probe AC-magnetic field with 0.5~Oe amplitude was used in the measurements. The results show a single sharp superconducting transition at $T_c^{\text{PR}}$ = 9.0~K for the pristine sample. However, a two-step transition is observed for GF01, revealing a second transition, related to the WL, at $T_c^{\text{WL}}$ = 8.7~K~\cite{valerio2021superconducting}.

Figure~\ref{Fig:Suppl}(f) depicts the variation of the field at which the magnetization peak is observed ($H_m$) as a function of temperature for GF01, GF02, and GF03. The $H_m$ values are obtained from $m(H)$ curves measured at different temperatures. It shows that, as temperature increases, the samples exposed to lower FIB doses still present a maximum in $m(H)$, whereas those exposed to higher doses do not. This is associated with a reduction of the ratio between the zero-field current able to flow through the WL in comparison to the PR, $J_{c0}^{WL}/J_{c0}$. As the films patterned by a higher FIB dose present a larger concentration of defects at the WL, they experience a further reduction of $J_{c0}^{WL}$. Therefore, as the emergence of the magnetization peak depends on $J_{c0}^{WL}/J_{c0}$, GF03 will not present the reported phenomenon at lower temperatures when compared to GF02, which, in turn, does not do so at lower temperatures than GF01.

\section{Magnetic flux penetration}
\label{Sc:AppB}

\begin{video}
\includegraphics[width=\linewidth]{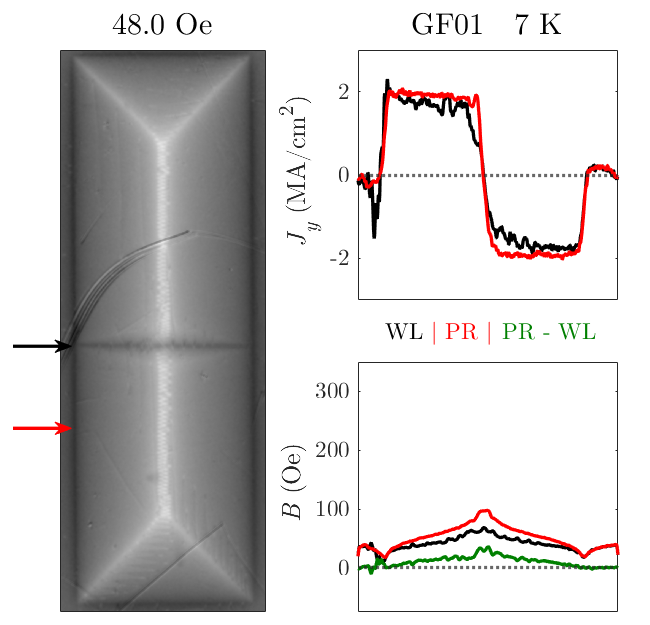}
\caption{\label{vid:GF01}The magnetic flux and shielding current behavior in GF01 at 7~K as the applied magnetic field is increased and decreased. The dark and red profiles highlight the behavior along the WL and the PR, respectively.}
\end{video}

Video~\ref{vid:GF01} shows MOI of GF01 at 47 different applied magnetic fields, both while $H$ is increased after ZFC to 7~K and while it is decreased from a maximum applied value of 315.6~Oe. Besides the MO images, $J_y$ and $B$ profiles are shown for each of the presented $H$ values. The profiles are evaluated on the PR (in red) and along the WL (in black), as indicated by color-matched arrows beside the MO images. A green profile is also presented and it is obtained by subtracting $B$ across the WL from $B$ across the PR. Noticeably, there is a scratch on the MO indicator film beside the left side of the groove. This creates rather visual discontinuities on the left side of the black profiles but does not interfere with the interpretation of the behavior inside the WL. The video corroborates the discussion presented in the main text by showing the gradual increase in $J_y$ across the WL as $H$ is reduced accompanied by the relative increase in $B$ in the same region. This happens as the central discontinuity lines, visible in the MO images, close toward the groove. 

Video~\ref{vid:GF01cross} shows the flux penetration patterns into sample GF01+ at 7~K after zero-field cooling as $H$ is increased up to 500~Oe and subsequently decreased back to 0~Oe. The video corroborates the simultaneous d-line movement along all four weak-links as the applied field is reduced, as discussed in the main text.

\begin{video}
\includegraphics[width=\linewidth]{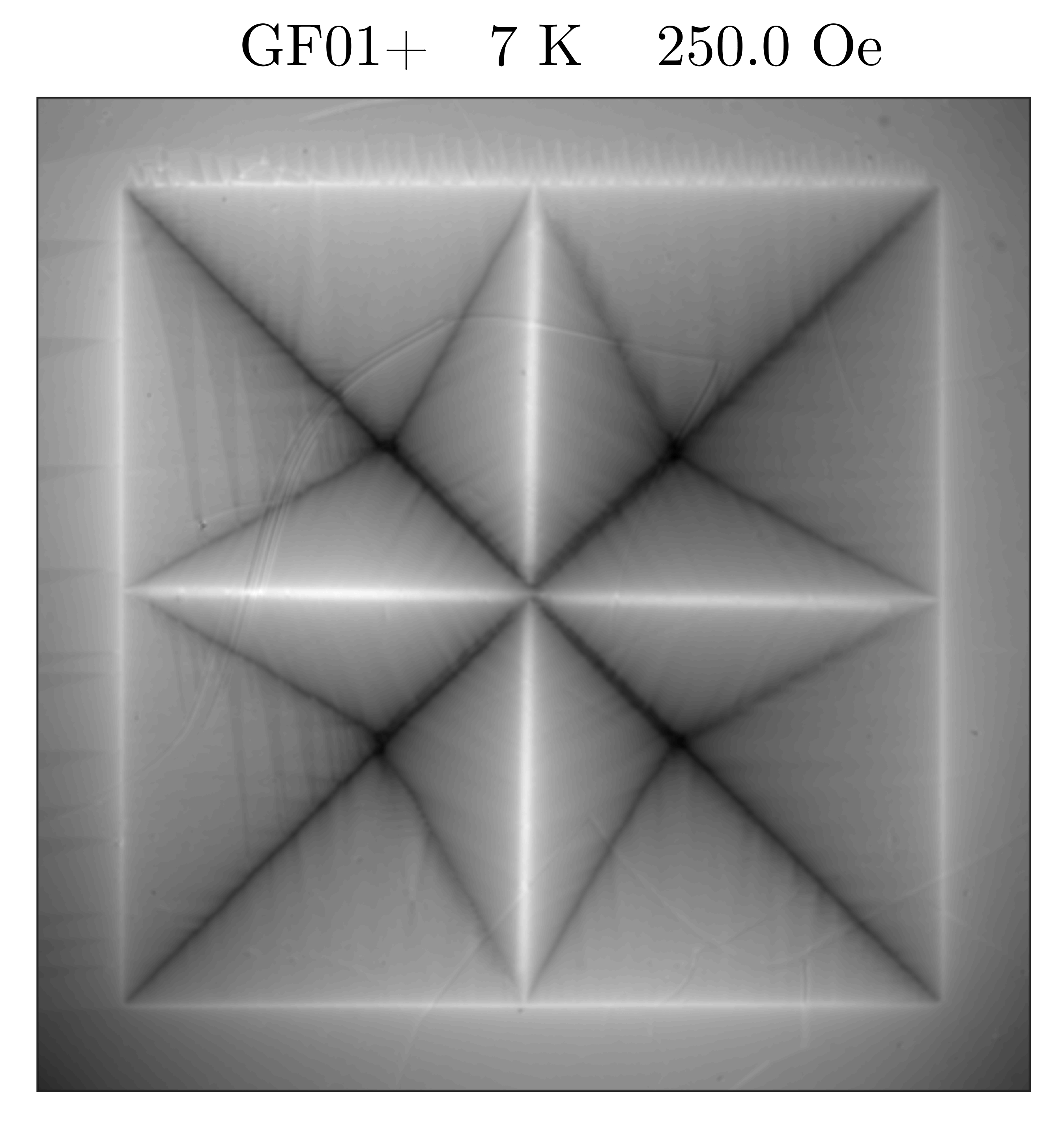}
\caption{\label{vid:GF01cross}Magnetic flux penetration into GF01$+$ at 7~K as the applied magnetic field is increased and decreased.}
\end{video}

\newpage

\bibliography{references}

\begin{thebibliography}{78}%
\makeatletter
\providecommand \@ifxundefined [1]{%
 \@ifx{#1\undefined}
}%
\providecommand \@ifnum [1]{%
 \ifnum #1\expandafter \@firstoftwo
 \else \expandafter \@secondoftwo
 \fi
}%
\providecommand \@ifx [1]{%
 \ifx #1\expandafter \@firstoftwo
 \else \expandafter \@secondoftwo
 \fi
}%
\providecommand \natexlab [1]{#1}%
\providecommand \enquote  [1]{``#1''}%
\providecommand \bibnamefont  [1]{#1}%
\providecommand \bibfnamefont [1]{#1}%
\providecommand \citenamefont [1]{#1}%
\providecommand \href@noop [0]{\@secondoftwo}%
\providecommand \href [0]{\begingroup \@sanitize@url \@href}%
\providecommand \@href[1]{\@@startlink{#1}\@@href}%
\providecommand \@@href[1]{\endgroup#1\@@endlink}%
\providecommand \@sanitize@url [0]{\catcode `\\12\catcode `\$12\catcode `\&12\catcode `\#12\catcode `\^12\catcode `\_12\catcode `\%12\relax}%
\providecommand \@@startlink[1]{}%
\providecommand \@@endlink[0]{}%
\providecommand \url  [0]{\begingroup\@sanitize@url \@url }%
\providecommand \@url [1]{\endgroup\@href {#1}{\urlprefix }}%
\providecommand \urlprefix  [0]{URL }%
\providecommand \Eprint [0]{\href }%
\providecommand \doibase [0]{https://doi.org/}%
\providecommand \selectlanguage [0]{\@gobble}%
\providecommand \bibinfo  [0]{\@secondoftwo}%
\providecommand \bibfield  [0]{\@secondoftwo}%
\providecommand \translation [1]{[#1]}%
\providecommand \BibitemOpen [0]{}%
\providecommand \bibitemStop [0]{}%
\providecommand \bibitemNoStop [0]{.\EOS\space}%
\providecommand \EOS [0]{\spacefactor3000\relax}%
\providecommand \BibitemShut  [1]{\csname bibitem#1\endcsname}%
\let\auto@bib@innerbib\@empty
\bibitem [{\citenamefont {Harada}\ \emph {et~al.}(1996)\citenamefont {Harada}, \citenamefont {Kamimura}, \citenamefont {Kasai}, \citenamefont {Matsuda}, \citenamefont {Tonomura},\ and\ \citenamefont {Moshchalkov}}]{harada1996direct}%
  \BibitemOpen
  \bibfield  {author} {\bibinfo {author} {\bibfnamefont {K.}~\bibnamefont {Harada}}, \bibinfo {author} {\bibfnamefont {O.}~\bibnamefont {Kamimura}}, \bibinfo {author} {\bibfnamefont {H.}~\bibnamefont {Kasai}}, \bibinfo {author} {\bibfnamefont {T.}~\bibnamefont {Matsuda}}, \bibinfo {author} {\bibfnamefont {A.}~\bibnamefont {Tonomura}},\ and\ \bibinfo {author} {\bibfnamefont {V.~V.}\ \bibnamefont {Moshchalkov}},\ }\bibfield  {title} {\bibinfo {title} {Direct observation of vortex dynamics in superconducting films with regular arrays of defects},\ }\href {https://doi.org/10.1126/science.274.5290.1167} {\bibfield  {journal} {\bibinfo  {journal} {Science}\ }\textbf {\bibinfo {volume} {274}},\ \bibinfo {pages} {1167} (\bibinfo {year} {1996})}\BibitemShut {NoStop}%
\bibitem [{\citenamefont {Silhanek}\ \emph {et~al.}(2003)\citenamefont {Silhanek}, \citenamefont {Van~Look}, \citenamefont {Raedts}, \citenamefont {Jonckheere},\ and\ \citenamefont {Moshchalkov}}]{silhanek2003guided}%
  \BibitemOpen
  \bibfield  {author} {\bibinfo {author} {\bibfnamefont {A.~V.}\ \bibnamefont {Silhanek}}, \bibinfo {author} {\bibfnamefont {L.}~\bibnamefont {Van~Look}}, \bibinfo {author} {\bibfnamefont {S.}~\bibnamefont {Raedts}}, \bibinfo {author} {\bibfnamefont {R.}~\bibnamefont {Jonckheere}},\ and\ \bibinfo {author} {\bibfnamefont {V.~V.}\ \bibnamefont {Moshchalkov}},\ }\bibfield  {title} {\bibinfo {title} {Guided vortex motion in superconductors with a square antidot array},\ }\href {https://doi.org/10.1103/PhysRevB.68.214504} {\bibfield  {journal} {\bibinfo  {journal} {Phys. Rev. B}\ }\textbf {\bibinfo {volume} {68}},\ \bibinfo {pages} {214504} (\bibinfo {year} {2003})}\BibitemShut {NoStop}%
\bibitem [{\citenamefont {Dobrovolskiy}\ \emph {et~al.}(2012)\citenamefont {Dobrovolskiy}, \citenamefont {Begun}, \citenamefont {Huth},\ and\ \citenamefont {Shklovskij}}]{dobrovolskiy2012electrical}%
  \BibitemOpen
  \bibfield  {author} {\bibinfo {author} {\bibfnamefont {O.~V.}\ \bibnamefont {Dobrovolskiy}}, \bibinfo {author} {\bibfnamefont {E.}~\bibnamefont {Begun}}, \bibinfo {author} {\bibfnamefont {M.}~\bibnamefont {Huth}},\ and\ \bibinfo {author} {\bibfnamefont {V.~A.}\ \bibnamefont {Shklovskij}},\ }\bibfield  {title} {\bibinfo {title} {Electrical transport and pinning properties of {Nb} thin films patterned with focused ion beam-milled washboard nanostructures},\ }\href {https://doi.org/10.1088/1367-2630/14/11/113027} {\bibfield  {journal} {\bibinfo  {journal} {New J. Phys.}\ }\textbf {\bibinfo {volume} {14}},\ \bibinfo {pages} {113027} (\bibinfo {year} {2012})}\BibitemShut {NoStop}%
\bibitem [{\citenamefont {Shaw}\ \emph {et~al.}(2012)\citenamefont {Shaw}, \citenamefont {Mandal}, \citenamefont {Bag}, \citenamefont {Banerjee}, \citenamefont {Tamegai},\ and\ \citenamefont {Suderow}}]{shaw2012properties}%
  \BibitemOpen
  \bibfield  {author} {\bibinfo {author} {\bibfnamefont {G.}~\bibnamefont {Shaw}}, \bibinfo {author} {\bibfnamefont {P.}~\bibnamefont {Mandal}}, \bibinfo {author} {\bibfnamefont {B.}~\bibnamefont {Bag}}, \bibinfo {author} {\bibfnamefont {S.}~\bibnamefont {Banerjee}}, \bibinfo {author} {\bibfnamefont {T.}~\bibnamefont {Tamegai}},\ and\ \bibinfo {author} {\bibfnamefont {H.}~\bibnamefont {Suderow}},\ }\bibfield  {title} {\bibinfo {title} {Properties of nanopatterned pins generated in a superconductor with {FIB}},\ }\href {https://doi.org/10.1016/j.apsusc.2011.05.011} {\bibfield  {journal} {\bibinfo  {journal} {Appl. Surf. Sci.}\ }\textbf {\bibinfo {volume} {258}},\ \bibinfo {pages} {4199} (\bibinfo {year} {2012})}\BibitemShut {NoStop}%
\bibitem [{\citenamefont {Wang}\ \emph {et~al.}(2013)\citenamefont {Wang}, \citenamefont {Latimer}, \citenamefont {Xiao}, \citenamefont {Divan}, \citenamefont {Ocola}, \citenamefont {Crabtree},\ and\ \citenamefont {Kwok}}]{wang2013enhancing}%
  \BibitemOpen
  \bibfield  {author} {\bibinfo {author} {\bibfnamefont {Y.~L.}\ \bibnamefont {Wang}}, \bibinfo {author} {\bibfnamefont {M.~L.}\ \bibnamefont {Latimer}}, \bibinfo {author} {\bibfnamefont {Z.~L.}\ \bibnamefont {Xiao}}, \bibinfo {author} {\bibfnamefont {R.}~\bibnamefont {Divan}}, \bibinfo {author} {\bibfnamefont {L.~E.}\ \bibnamefont {Ocola}}, \bibinfo {author} {\bibfnamefont {G.~W.}\ \bibnamefont {Crabtree}},\ and\ \bibinfo {author} {\bibfnamefont {W.~K.}\ \bibnamefont {Kwok}},\ }\bibfield  {title} {\bibinfo {title} {Enhancing the critical current of a superconducting film in a wide range of magnetic fields with a conformal array of nanoscale holes},\ }\href {https://doi.org/10.1103/PhysRevB.87.220501} {\bibfield  {journal} {\bibinfo  {journal} {Phys. Rev. B}\ }\textbf {\bibinfo {volume} {87}},\ \bibinfo {pages} {220501} (\bibinfo {year} {2013})}\BibitemShut {NoStop}%
\bibitem [{\citenamefont {Dobrovolskiy}(2017)}]{dobrovolskiy2017abrikosov}%
  \BibitemOpen
  \bibfield  {author} {\bibinfo {author} {\bibfnamefont {O.~V.}\ \bibnamefont {Dobrovolskiy}},\ }\bibfield  {title} {\bibinfo {title} {Abrikosov fluxonics in washboard nanolandscapes},\ }\href {https://doi.org/https://doi.org/10.1016/j.physc.2016.07.008} {\bibfield  {journal} {\bibinfo  {journal} {Physica C}\ }\textbf {\bibinfo {volume} {533}},\ \bibinfo {pages} {80} (\bibinfo {year} {2017})}\BibitemShut {NoStop}%
\bibitem [{\citenamefont {Kalcheim}\ \emph {et~al.}(2017)\citenamefont {Kalcheim}, \citenamefont {Katzir}, \citenamefont {Zeides}, \citenamefont {Katz}, \citenamefont {Paltiel},\ and\ \citenamefont {Millo}}]{kalcheim2017dynamic}%
  \BibitemOpen
  \bibfield  {author} {\bibinfo {author} {\bibfnamefont {Y.}~\bibnamefont {Kalcheim}}, \bibinfo {author} {\bibfnamefont {E.}~\bibnamefont {Katzir}}, \bibinfo {author} {\bibfnamefont {F.}~\bibnamefont {Zeides}}, \bibinfo {author} {\bibfnamefont {N.}~\bibnamefont {Katz}}, \bibinfo {author} {\bibfnamefont {Y.}~\bibnamefont {Paltiel}},\ and\ \bibinfo {author} {\bibfnamefont {O.}~\bibnamefont {Millo}},\ }\bibfield  {title} {\bibinfo {title} {Dynamic control of the vortex pinning potential in a superconductor using current injection through nanoscale patterns},\ }\href {https://doi.org/10.1021/acs.nanolett.7b00179} {\bibfield  {journal} {\bibinfo  {journal} {Nano Lett.}\ }\textbf {\bibinfo {volume} {17}},\ \bibinfo {pages} {2934} (\bibinfo {year} {2017})}\BibitemShut {NoStop}%
\bibitem [{\citenamefont {Dobrovolskiy}\ \emph {et~al.}(2017)\citenamefont {Dobrovolskiy}, \citenamefont {Shklovskij}, \citenamefont {Hanefeld}, \citenamefont {Zörb}, \citenamefont {Köhs},\ and\ \citenamefont {Huth}}]{dobrovolskiy2017pinning}%
  \BibitemOpen
  \bibfield  {author} {\bibinfo {author} {\bibfnamefont {O.~V.}\ \bibnamefont {Dobrovolskiy}}, \bibinfo {author} {\bibfnamefont {V.~A.}\ \bibnamefont {Shklovskij}}, \bibinfo {author} {\bibfnamefont {M.}~\bibnamefont {Hanefeld}}, \bibinfo {author} {\bibfnamefont {M.}~\bibnamefont {Zörb}}, \bibinfo {author} {\bibfnamefont {L.}~\bibnamefont {Köhs}},\ and\ \bibinfo {author} {\bibfnamefont {M.}~\bibnamefont {Huth}},\ }\bibfield  {title} {\bibinfo {title} {Pinning effects on flux flow instability in epitaxial {Nb} thin films},\ }\href {https://doi.org/10.1088/1361-6668/aa73aa} {\bibfield  {journal} {\bibinfo  {journal} {Supercond. Sci. Technol.}\ }\textbf {\bibinfo {volume} {30}},\ \bibinfo {pages} {085002} (\bibinfo {year} {2017})}\BibitemShut {NoStop}%
\bibitem [{\citenamefont {Colauto}\ \emph {et~al.}(2020)\citenamefont {Colauto}, \citenamefont {Motta},\ and\ \citenamefont {Ortiz}}]{colauto2021controlling}%
  \BibitemOpen
  \bibfield  {author} {\bibinfo {author} {\bibfnamefont {F.}~\bibnamefont {Colauto}}, \bibinfo {author} {\bibfnamefont {M.}~\bibnamefont {Motta}},\ and\ \bibinfo {author} {\bibfnamefont {W.~A.}\ \bibnamefont {Ortiz}},\ }\bibfield  {title} {\bibinfo {title} {Controlling magnetic flux penetration in {low-$T_c$} superconducting films and hybrids},\ }\href {https://doi.org/10.1088/1361-6668/abac1e} {\bibfield  {journal} {\bibinfo  {journal} {Supercond. Sci. Technol.}\ }\textbf {\bibinfo {volume} {34}},\ \bibinfo {pages} {013002} (\bibinfo {year} {2020})}\BibitemShut {NoStop}%
\bibitem [{\citenamefont {Likharev}(1979)}]{likharev1979superconducting}%
  \BibitemOpen
  \bibfield  {author} {\bibinfo {author} {\bibfnamefont {K.~K.}\ \bibnamefont {Likharev}},\ }\bibfield  {title} {\bibinfo {title} {Superconducting weak links},\ }\href {https://doi.org/10.1103/RevModPhys.51.101} {\bibfield  {journal} {\bibinfo  {journal} {Rev. Mod. Phys.}\ }\textbf {\bibinfo {volume} {51}},\ \bibinfo {pages} {101} (\bibinfo {year} {1979})}\BibitemShut {NoStop}%
\bibitem [{\citenamefont {Moshchalkov}\ \emph {et~al.}(2010)\citenamefont {Moshchalkov}, \citenamefont {Woerdenweber},\ and\ \citenamefont {Lang}}]{moshchalkov2010nanoscience}%
  \BibitemOpen
  \bibinfo {editor} {\bibfnamefont {V.}~\bibnamefont {Moshchalkov}}, \bibinfo {editor} {\bibfnamefont {R.}~\bibnamefont {Woerdenweber}},\ and\ \bibinfo {editor} {\bibfnamefont {W.}~\bibnamefont {Lang}},\ eds.,\ \href {https://doi.org/10.1007/978-3-642-15137-8} {\emph {\bibinfo {title} {Nanoscience and engineering in superconductivity}}}\ (\bibinfo  {publisher} {Springer Berlin, Heidelberg},\ \bibinfo {year} {2010})\BibitemShut {NoStop}%
\bibitem [{\citenamefont {Fornieri}\ \emph {et~al.}(2019)\citenamefont {Fornieri}, \citenamefont {Whiticar}, \citenamefont {Setiawan}, \citenamefont {Portol{\'e}s}, \citenamefont {Drachmann}, \citenamefont {Keselman}, \citenamefont {Gronin}, \citenamefont {Thomas}, \citenamefont {Wang}, \citenamefont {Kallaher} \emph {et~al.}}]{fornieri2019evidence}%
  \BibitemOpen
  \bibfield  {author} {\bibinfo {author} {\bibfnamefont {A.}~\bibnamefont {Fornieri}}, \bibinfo {author} {\bibfnamefont {A.~M.}\ \bibnamefont {Whiticar}}, \bibinfo {author} {\bibfnamefont {F.}~\bibnamefont {Setiawan}}, \bibinfo {author} {\bibfnamefont {E.}~\bibnamefont {Portol{\'e}s}}, \bibinfo {author} {\bibfnamefont {A.~C.}\ \bibnamefont {Drachmann}}, \bibinfo {author} {\bibfnamefont {A.}~\bibnamefont {Keselman}}, \bibinfo {author} {\bibfnamefont {S.}~\bibnamefont {Gronin}}, \bibinfo {author} {\bibfnamefont {C.}~\bibnamefont {Thomas}}, \bibinfo {author} {\bibfnamefont {T.}~\bibnamefont {Wang}}, \bibinfo {author} {\bibfnamefont {R.}~\bibnamefont {Kallaher}}, \emph {et~al.},\ }\bibfield  {title} {\bibinfo {title} {Evidence of topological superconductivity in planar {J}osephson junctions},\ }\href {https://doi.org/10.1038/s41586-019-1068-8} {\bibfield  {journal} {\bibinfo  {journal} {Nature}\ }\textbf {\bibinfo {volume} {569}},\ \bibinfo {pages} {89} (\bibinfo {year} {2019})}\BibitemShut {NoStop}%
\bibitem [{\citenamefont {Holzman}\ and\ \citenamefont {Ivry}(2019)}]{holzman2019superconducting}%
  \BibitemOpen
  \bibfield  {author} {\bibinfo {author} {\bibfnamefont {I.}~\bibnamefont {Holzman}}\ and\ \bibinfo {author} {\bibfnamefont {Y.}~\bibnamefont {Ivry}},\ }\bibfield  {title} {\bibinfo {title} {Superconducting nanowires for single-photon detection: Progress, challenges, and opportunities},\ }\href {https://doi.org/10.1002/qute.201800058} {\bibfield  {journal} {\bibinfo  {journal} {Adv. Quantum Technol.}\ }\textbf {\bibinfo {volume} {2}},\ \bibinfo {pages} {1800058} (\bibinfo {year} {2019})}\BibitemShut {NoStop}%
\bibitem [{\citenamefont {Murphy}\ \emph {et~al.}(2017)\citenamefont {Murphy}, \citenamefont {Averin},\ and\ \citenamefont {Bezryadin}}]{murphy2017nanoscale}%
  \BibitemOpen
  \bibfield  {author} {\bibinfo {author} {\bibfnamefont {A.}~\bibnamefont {Murphy}}, \bibinfo {author} {\bibfnamefont {D.~V.}\ \bibnamefont {Averin}},\ and\ \bibinfo {author} {\bibfnamefont {A.}~\bibnamefont {Bezryadin}},\ }\bibfield  {title} {\bibinfo {title} {Nanoscale superconducting memory based on the kinetic inductance of asymmetric nanowire loops},\ }\href {https://doi.org/10.1088/1367-2630/aa7331} {\bibfield  {journal} {\bibinfo  {journal} {New J. Phys.}\ }\textbf {\bibinfo {volume} {19}},\ \bibinfo {pages} {063015} (\bibinfo {year} {2017})}\BibitemShut {NoStop}%
\bibitem [{\citenamefont {Chen}\ \emph {et~al.}(2020)\citenamefont {Chen} \emph {et~al.}}]{chen2020miniaturization}%
  \BibitemOpen
  \bibfield  {author} {\bibinfo {author} {\bibfnamefont {L.}~\bibnamefont {Chen}} \emph {et~al.},\ }\bibfield  {title} {\bibinfo {title} {Miniaturization of the superconducting memory cell via a three-dimensional {Nb} nano-superconducting quantum interference device},\ }\href {https://doi.org/10.1021/acsnano.0c04405} {\bibfield  {journal} {\bibinfo  {journal} {ACS Nano}\ }\textbf {\bibinfo {volume} {14}},\ \bibinfo {pages} {11002} (\bibinfo {year} {2020})}\BibitemShut {NoStop}%
\bibitem [{\citenamefont {Ilin}\ \emph {et~al.}(2021)\citenamefont {Ilin}, \citenamefont {Song}, \citenamefont {Burkova}, \citenamefont {Silge}, \citenamefont {Guo}, \citenamefont {Ilin},\ and\ \citenamefont {Bezryadin}}]{ilin2021supercurrent}%
  \BibitemOpen
  \bibfield  {author} {\bibinfo {author} {\bibfnamefont {E.}~\bibnamefont {Ilin}}, \bibinfo {author} {\bibfnamefont {X.}~\bibnamefont {Song}}, \bibinfo {author} {\bibfnamefont {I.}~\bibnamefont {Burkova}}, \bibinfo {author} {\bibfnamefont {A.}~\bibnamefont {Silge}}, \bibinfo {author} {\bibfnamefont {Z.}~\bibnamefont {Guo}}, \bibinfo {author} {\bibfnamefont {K.}~\bibnamefont {Ilin}},\ and\ \bibinfo {author} {\bibfnamefont {A.}~\bibnamefont {Bezryadin}},\ }\bibfield  {title} {\bibinfo {title} {Supercurrent-controlled kinetic inductance superconducting memory element},\ }\href {https://doi.org/10.1063/5.0040563} {\bibfield  {journal} {\bibinfo  {journal} {Appl. Phys. Lett.}\ }\textbf {\bibinfo {volume} {118}},\ \bibinfo {pages} {112603} (\bibinfo {year} {2021})}\BibitemShut {NoStop}%
\bibitem [{\citenamefont {Chaves}\ \emph {et~al.}(2023)\citenamefont {Chaves}, \citenamefont {Nulens}, \citenamefont {Dausy}, \citenamefont {Raes}, \citenamefont {Yue}, \citenamefont {Ortiz}, \citenamefont {Motta}, \citenamefont {Van~Bael},\ and\ \citenamefont {Van~de Vondel}}]{chaves2023nanobridge}%
  \BibitemOpen
  \bibfield  {author} {\bibinfo {author} {\bibfnamefont {D.~A.~D.}\ \bibnamefont {Chaves}}, \bibinfo {author} {\bibfnamefont {L.}~\bibnamefont {Nulens}}, \bibinfo {author} {\bibfnamefont {H.}~\bibnamefont {Dausy}}, \bibinfo {author} {\bibfnamefont {B.}~\bibnamefont {Raes}}, \bibinfo {author} {\bibfnamefont {D.}~\bibnamefont {Yue}}, \bibinfo {author} {\bibfnamefont {W.~A.}\ \bibnamefont {Ortiz}}, \bibinfo {author} {\bibfnamefont {M.}~\bibnamefont {Motta}}, \bibinfo {author} {\bibfnamefont {M.~J.}\ \bibnamefont {Van~Bael}},\ and\ \bibinfo {author} {\bibfnamefont {J.}~\bibnamefont {Van~de Vondel}},\ }\bibfield  {title} {\bibinfo {title} {Nanobridge {SQUIDs} as multilevel memory elements},\ }\href {https://doi.org/10.1103/PhysRevApplied.19.034091} {\bibfield  {journal} {\bibinfo  {journal} {Phys. Rev. Appl.}\ }\textbf {\bibinfo {volume} {19}},\ \bibinfo {pages} {034091} (\bibinfo {year} {2023})}\BibitemShut {NoStop}%
\bibitem [{\citenamefont {Lyu}\ \emph {et~al.}(2021)\citenamefont {Lyu} \emph {et~al.}}]{lyu2021superconducting}%
  \BibitemOpen
  \bibfield  {author} {\bibinfo {author} {\bibfnamefont {Y.-Y.}\ \bibnamefont {Lyu}} \emph {et~al.},\ }\bibfield  {title} {\bibinfo {title} {Superconducting diode effect via conformal-mapped nanoholes},\ }\href {https://doi.org/10.1038/s41467-021-23077-0} {\bibfield  {journal} {\bibinfo  {journal} {Nat. Commun.}\ }\textbf {\bibinfo {volume} {12}},\ \bibinfo {pages} {1} (\bibinfo {year} {2021})}\BibitemShut {NoStop}%
\bibitem [{\citenamefont {Golod}\ and\ \citenamefont {Krasnov}(2022)}]{golod2022demonstration}%
  \BibitemOpen
  \bibfield  {author} {\bibinfo {author} {\bibfnamefont {T.}~\bibnamefont {Golod}}\ and\ \bibinfo {author} {\bibfnamefont {V.~M.}\ \bibnamefont {Krasnov}},\ }\bibfield  {title} {\bibinfo {title} {Demonstration of a superconducting diode-with-memory, operational at zero magnetic field with switchable nonreciprocity},\ }\href {https://doi.org/10.1038/s41467-022-31256-w} {\bibfield  {journal} {\bibinfo  {journal} {Nat. Commun.}\ }\textbf {\bibinfo {volume} {13}},\ \bibinfo {pages} {3658} (\bibinfo {year} {2022})}\BibitemShut {NoStop}%
\bibitem [{\citenamefont {Strambini}\ \emph {et~al.}(2020)\citenamefont {Strambini} \emph {et~al.}}]{strambini2020josephson}%
  \BibitemOpen
  \bibfield  {author} {\bibinfo {author} {\bibfnamefont {E.}~\bibnamefont {Strambini}} \emph {et~al.},\ }\bibfield  {title} {\bibinfo {title} {A {J}osephson phase battery},\ }\href {https://doi.org/10.1038/s41565-020-0712-7} {\bibfield  {journal} {\bibinfo  {journal} {Nat. Nanotechnol.}\ }\textbf {\bibinfo {volume} {15}},\ \bibinfo {pages} {656} (\bibinfo {year} {2020})}\BibitemShut {NoStop}%
\bibitem [{\citenamefont {Yamashita}\ \emph {et~al.}(2020)\citenamefont {Yamashita}, \citenamefont {Kim}, \citenamefont {Kato}, \citenamefont {Qiu}, \citenamefont {Semba}, \citenamefont {Fujimaki},\ and\ \citenamefont {Terai}}]{yamashita2020pi}%
  \BibitemOpen
  \bibfield  {author} {\bibinfo {author} {\bibfnamefont {T.}~\bibnamefont {Yamashita}}, \bibinfo {author} {\bibfnamefont {S.}~\bibnamefont {Kim}}, \bibinfo {author} {\bibfnamefont {H.}~\bibnamefont {Kato}}, \bibinfo {author} {\bibfnamefont {W.}~\bibnamefont {Qiu}}, \bibinfo {author} {\bibfnamefont {K.}~\bibnamefont {Semba}}, \bibinfo {author} {\bibfnamefont {A.}~\bibnamefont {Fujimaki}},\ and\ \bibinfo {author} {\bibfnamefont {H.}~\bibnamefont {Terai}},\ }\bibfield  {title} {\bibinfo {title} {$\pi$ phase shifter based on {NbN}-based ferromagnetic {Josephson} junction on a silicon substrate},\ }\href {https://doi.org/10.1038/s41598-020-70766-9} {\bibfield  {journal} {\bibinfo  {journal} {Sci. Rep.}\ }\textbf {\bibinfo {volume} {10}},\ \bibinfo {pages} {1} (\bibinfo {year} {2020})}\BibitemShut {NoStop}%
\bibitem [{\citenamefont {Golod}\ \emph {et~al.}(2021)\citenamefont {Golod}, \citenamefont {Hovhannisyan}, \citenamefont {Kapran}, \citenamefont {Dremov}, \citenamefont {Stolyarov},\ and\ \citenamefont {Krasnov}}]{golod2021reconFigurable}%
  \BibitemOpen
  \bibfield  {author} {\bibinfo {author} {\bibfnamefont {T.}~\bibnamefont {Golod}}, \bibinfo {author} {\bibfnamefont {R.~A.}\ \bibnamefont {Hovhannisyan}}, \bibinfo {author} {\bibfnamefont {O.~M.}\ \bibnamefont {Kapran}}, \bibinfo {author} {\bibfnamefont {V.~V.}\ \bibnamefont {Dremov}}, \bibinfo {author} {\bibfnamefont {V.~S.}\ \bibnamefont {Stolyarov}},\ and\ \bibinfo {author} {\bibfnamefont {V.~M.}\ \bibnamefont {Krasnov}},\ }\bibfield  {title} {\bibinfo {title} {Reconfigurable {J}osephson phase shifter},\ }\href {https://doi.org/10.1021/acs.nanolett.1c01366} {\bibfield  {journal} {\bibinfo  {journal} {Nano Lett.}\ }\textbf {\bibinfo {volume} {21}},\ \bibinfo {pages} {5240} (\bibinfo {year} {2021})}\BibitemShut {NoStop}%
\bibitem [{\citenamefont {Volkert}\ and\ \citenamefont {Minor}(2007)}]{volkert2007focused}%
  \BibitemOpen
  \bibfield  {author} {\bibinfo {author} {\bibfnamefont {C.~A.}\ \bibnamefont {Volkert}}\ and\ \bibinfo {author} {\bibfnamefont {A.~M.}\ \bibnamefont {Minor}},\ }\bibfield  {title} {\bibinfo {title} {Focused ion beam microscopy and micromachining},\ }\href {https://doi.org/10.1557/mrs2007.62} {\bibfield  {journal} {\bibinfo  {journal} {MRS Bull.}\ }\textbf {\bibinfo {volume} {32}},\ \bibinfo {pages} {389} (\bibinfo {year} {2007})}\BibitemShut {NoStop}%
\bibitem [{\citenamefont {Wyss}\ \emph {et~al.}(2022)\citenamefont {Wyss}, \citenamefont {Bagani}, \citenamefont {Jetter}, \citenamefont {Marchiori}, \citenamefont {Vervelaki}, \citenamefont {Gross}, \citenamefont {Ridderbos}, \citenamefont {Gliga}, \citenamefont {Sch\"onenberger},\ and\ \citenamefont {Poggio}}]{wyss2022magnetic}%
  \BibitemOpen
  \bibfield  {author} {\bibinfo {author} {\bibfnamefont {M.}~\bibnamefont {Wyss}}, \bibinfo {author} {\bibfnamefont {K.}~\bibnamefont {Bagani}}, \bibinfo {author} {\bibfnamefont {D.}~\bibnamefont {Jetter}}, \bibinfo {author} {\bibfnamefont {E.}~\bibnamefont {Marchiori}}, \bibinfo {author} {\bibfnamefont {A.}~\bibnamefont {Vervelaki}}, \bibinfo {author} {\bibfnamefont {B.}~\bibnamefont {Gross}}, \bibinfo {author} {\bibfnamefont {J.}~\bibnamefont {Ridderbos}}, \bibinfo {author} {\bibfnamefont {S.}~\bibnamefont {Gliga}}, \bibinfo {author} {\bibfnamefont {C.}~\bibnamefont {Sch\"onenberger}},\ and\ \bibinfo {author} {\bibfnamefont {M.}~\bibnamefont {Poggio}},\ }\bibfield  {title} {\bibinfo {title} {Magnetic, thermal, and topographic imaging with a nanometer-scale {SQUID}-on-lever scanning probe},\ }\href {https://doi.org/10.1103/PhysRevApplied.17.034002} {\bibfield  {journal} {\bibinfo  {journal} {Phys. Rev. Appl.}\ }\textbf {\bibinfo {volume} {17}},\ \bibinfo {pages} {034002} (\bibinfo {year} {2022})}\BibitemShut
  {NoStop}%
\bibitem [{\citenamefont {Sigloch}\ \emph {et~al.}(2022)\citenamefont {Sigloch}, \citenamefont {Sangiao}, \citenamefont {Orús},\ and\ \citenamefont {de~Teresa}}]{sigloch2022direct}%
  \BibitemOpen
  \bibfield  {author} {\bibinfo {author} {\bibfnamefont {F.}~\bibnamefont {Sigloch}}, \bibinfo {author} {\bibfnamefont {S.}~\bibnamefont {Sangiao}}, \bibinfo {author} {\bibfnamefont {P.}~\bibnamefont {Orús}},\ and\ \bibinfo {author} {\bibfnamefont {J.~M.}\ \bibnamefont {de~Teresa}},\ }\bibfield  {title} {\bibinfo {title} {Direct-write of tungsten-carbide {nanoSQUIDs} based on focused ion beam induced deposition},\ }\href {https://doi.org/10.1039/D2NA00602B} {\bibfield  {journal} {\bibinfo  {journal} {Nanoscale Adv.}\ }\textbf {\bibinfo {volume} {4}},\ \bibinfo {pages} {4628} (\bibinfo {year} {2022})}\BibitemShut {NoStop}%
\bibitem [{\citenamefont {Datesman}\ \emph {et~al.}(2005)\citenamefont {Datesman}, \citenamefont {Schultz}, \citenamefont {Cecil}, \citenamefont {Lyons},\ and\ \citenamefont {Lichtenberger}}]{datesman2005gallium}%
  \BibitemOpen
  \bibfield  {author} {\bibinfo {author} {\bibfnamefont {A.~M.}\ \bibnamefont {Datesman}}, \bibinfo {author} {\bibfnamefont {J.~C.}\ \bibnamefont {Schultz}}, \bibinfo {author} {\bibfnamefont {T.~W.}\ \bibnamefont {Cecil}}, \bibinfo {author} {\bibfnamefont {C.~M.}\ \bibnamefont {Lyons}},\ and\ \bibinfo {author} {\bibfnamefont {A.~W.}\ \bibnamefont {Lichtenberger}},\ }\bibfield  {title} {\bibinfo {title} {Gallium ion implantation into niobium thin films using a focused-ion beam},\ }\href {https://doi.org/10.1109/TASC.2005.849029} {\bibfield  {journal} {\bibinfo  {journal} {IEEE Trans. Appl. Supercond.}\ }\textbf {\bibinfo {volume} {15}},\ \bibinfo {pages} {3524} (\bibinfo {year} {2005})}\BibitemShut {NoStop}%
\bibitem [{\citenamefont {De~Leo}\ \emph {et~al.}(2016)\citenamefont {De~Leo}, \citenamefont {Fretto}, \citenamefont {Lacquaniti}, \citenamefont {Cassiago}, \citenamefont {D'Ortenzi}, \citenamefont {Boarino},\ and\ \citenamefont {Maggi}}]{deleo2016thickness}%
  \BibitemOpen
  \bibfield  {author} {\bibinfo {author} {\bibfnamefont {N.}~\bibnamefont {De~Leo}}, \bibinfo {author} {\bibfnamefont {M.}~\bibnamefont {Fretto}}, \bibinfo {author} {\bibfnamefont {V.}~\bibnamefont {Lacquaniti}}, \bibinfo {author} {\bibfnamefont {C.}~\bibnamefont {Cassiago}}, \bibinfo {author} {\bibfnamefont {L.}~\bibnamefont {D'Ortenzi}}, \bibinfo {author} {\bibfnamefont {L.}~\bibnamefont {Boarino}},\ and\ \bibinfo {author} {\bibfnamefont {S.}~\bibnamefont {Maggi}},\ }\bibfield  {title} {\bibinfo {title} {Thickness modulated niobium nanoconstrictions by focused ion beam and anodization},\ }\href {https://doi.org/10.1109/TASC.2016.2542286} {\bibfield  {journal} {\bibinfo  {journal} {IEEE Trans. Appl. Supercond.}\ }\textbf {\bibinfo {volume} {26}},\ \bibinfo {pages} {1} (\bibinfo {year} {2016})}\BibitemShut {NoStop}%
\bibitem [{\citenamefont {Singh}\ \emph {et~al.}(2018)\citenamefont {Singh}, \citenamefont {Chaujar}, \citenamefont {Husale}, \citenamefont {Grover}, \citenamefont {Shah}, \citenamefont {Deshmukh}, \citenamefont {Gupta}, \citenamefont {Singh}, \citenamefont {Ojha}, \citenamefont {Aswal},\ and\ \citenamefont {Rakshit}}]{singh2018influence}%
  \BibitemOpen
  \bibfield  {author} {\bibinfo {author} {\bibfnamefont {M.}~\bibnamefont {Singh}}, \bibinfo {author} {\bibfnamefont {R.}~\bibnamefont {Chaujar}}, \bibinfo {author} {\bibfnamefont {S.}~\bibnamefont {Husale}}, \bibinfo {author} {\bibfnamefont {S.}~\bibnamefont {Grover}}, \bibinfo {author} {\bibfnamefont {A.~P.}\ \bibnamefont {Shah}}, \bibinfo {author} {\bibfnamefont {M.~M.}\ \bibnamefont {Deshmukh}}, \bibinfo {author} {\bibfnamefont {A.}~\bibnamefont {Gupta}}, \bibinfo {author} {\bibfnamefont {V.~N.}\ \bibnamefont {Singh}}, \bibinfo {author} {\bibfnamefont {V.~N.}\ \bibnamefont {Ojha}}, \bibinfo {author} {\bibfnamefont {D.~K.}\ \bibnamefont {Aswal}},\ and\ \bibinfo {author} {\bibfnamefont {R.~K.}\ \bibnamefont {Rakshit}},\ }\bibfield  {title} {\bibinfo {title} {Influence of fabrication processes on transport properties of superconducting niobium nitride nanowires},\ }\href@noop {} {\bibfield  {journal} {\bibinfo  {journal} {Current Science}\ }\textbf {\bibinfo {volume} {114}},\ \bibinfo {pages} {1443}
  (\bibinfo {year} {2018})}\BibitemShut {NoStop}%
\bibitem [{\citenamefont {Aichner}\ \emph {et~al.}(2019)\citenamefont {Aichner}, \citenamefont {Müller}, \citenamefont {Karrer}, \citenamefont {Misko}, \citenamefont {Limberger}, \citenamefont {Mletschnig}, \citenamefont {Dosmailov}, \citenamefont {Pedarnig}, \citenamefont {Nori}, \citenamefont {Kleiner}, \citenamefont {Koelle},\ and\ \citenamefont {Lang}}]{aichner2019ultradense}%
  \BibitemOpen
  \bibfield  {author} {\bibinfo {author} {\bibfnamefont {B.}~\bibnamefont {Aichner}}, \bibinfo {author} {\bibfnamefont {B.}~\bibnamefont {Müller}}, \bibinfo {author} {\bibfnamefont {M.}~\bibnamefont {Karrer}}, \bibinfo {author} {\bibfnamefont {V.~R.}\ \bibnamefont {Misko}}, \bibinfo {author} {\bibfnamefont {F.}~\bibnamefont {Limberger}}, \bibinfo {author} {\bibfnamefont {K.~L.}\ \bibnamefont {Mletschnig}}, \bibinfo {author} {\bibfnamefont {M.}~\bibnamefont {Dosmailov}}, \bibinfo {author} {\bibfnamefont {J.~D.}\ \bibnamefont {Pedarnig}}, \bibinfo {author} {\bibfnamefont {F.}~\bibnamefont {Nori}}, \bibinfo {author} {\bibfnamefont {R.}~\bibnamefont {Kleiner}}, \bibinfo {author} {\bibfnamefont {D.}~\bibnamefont {Koelle}},\ and\ \bibinfo {author} {\bibfnamefont {W.}~\bibnamefont {Lang}},\ }\bibfield  {title} {\bibinfo {title} {Ultradense tailored vortex pinning arrays in superconducting {YBa$_2$Cu$_3$O$_{7-\delta}$} thin films created by focused {He} ion beam irradiation for fluxonics applications},\ }\href
  {https://doi.org/10.1021/acsanm.9b01006} {\bibfield  {journal} {\bibinfo  {journal} {ACS Appl. Nano Mater.}\ }\textbf {\bibinfo {volume} {2}},\ \bibinfo {pages} {5108} (\bibinfo {year} {2019})}\BibitemShut {NoStop}%
\bibitem [{\citenamefont {Valerio-Cuadros}\ \emph {et~al.}(2021)\citenamefont {Valerio-Cuadros}, \citenamefont {Chaves}, \citenamefont {Colauto}, \citenamefont {Oliveira}, \citenamefont {Andrade}, \citenamefont {Johansen}, \citenamefont {Ortiz},\ and\ \citenamefont {Motta}}]{valerio2021superconducting}%
  \BibitemOpen
  \bibfield  {author} {\bibinfo {author} {\bibfnamefont {M.~I.}\ \bibnamefont {Valerio-Cuadros}}, \bibinfo {author} {\bibfnamefont {D.~A.~D.}\ \bibnamefont {Chaves}}, \bibinfo {author} {\bibfnamefont {F.}~\bibnamefont {Colauto}}, \bibinfo {author} {\bibfnamefont {A.~A.~M.}\ \bibnamefont {Oliveira}}, \bibinfo {author} {\bibfnamefont {A.~M.~H.}\ \bibnamefont {Andrade}}, \bibinfo {author} {\bibfnamefont {T.~H.}\ \bibnamefont {Johansen}}, \bibinfo {author} {\bibfnamefont {W.~A.}\ \bibnamefont {Ortiz}},\ and\ \bibinfo {author} {\bibfnamefont {M.}~\bibnamefont {Motta}},\ }\bibfield  {title} {\bibinfo {title} {Superconducting properties and electron scattering mechanisms in a {Nb} film with a single weak-link excavated by focused ion beam},\ }\href {https://doi.org/10.3390/ma14237274} {\bibfield  {journal} {\bibinfo  {journal} {Materials}\ }\textbf {\bibinfo {volume} {14}},\ \bibinfo {pages} {7274} (\bibinfo {year} {2021})}\BibitemShut {NoStop}%
\bibitem [{\citenamefont {Larbalestier}\ \emph {et~al.}(2001)\citenamefont {Larbalestier}, \citenamefont {Gurevich}, \citenamefont {Feldmann},\ and\ \citenamefont {Polyanskii}}]{larbalestier2001high}%
  \BibitemOpen
  \bibfield  {author} {\bibinfo {author} {\bibfnamefont {D.}~\bibnamefont {Larbalestier}}, \bibinfo {author} {\bibfnamefont {A.}~\bibnamefont {Gurevich}}, \bibinfo {author} {\bibfnamefont {D.~M.}\ \bibnamefont {Feldmann}},\ and\ \bibinfo {author} {\bibfnamefont {A.}~\bibnamefont {Polyanskii}},\ }\bibfield  {title} {\bibinfo {title} {{High-$T_c$} superconducting materials for electric power applications},\ }\href {https://doi.org/10.1038/35104654} {\bibfield  {journal} {\bibinfo  {journal} {Nature}\ }\textbf {\bibinfo {volume} {414}},\ \bibinfo {pages} {368} (\bibinfo {year} {2001})}\BibitemShut {NoStop}%
\bibitem [{\citenamefont {Foltyn}\ \emph {et~al.}(2007)\citenamefont {Foltyn}, \citenamefont {Civale}, \citenamefont {MacManus-Driscoll}, \citenamefont {Jia}, \citenamefont {Maiorov}, \citenamefont {Wang},\ and\ \citenamefont {Maley}}]{foltyn2007materials}%
  \BibitemOpen
  \bibfield  {author} {\bibinfo {author} {\bibfnamefont {S.~R.}\ \bibnamefont {Foltyn}}, \bibinfo {author} {\bibfnamefont {L.}~\bibnamefont {Civale}}, \bibinfo {author} {\bibfnamefont {J.~L.}\ \bibnamefont {MacManus-Driscoll}}, \bibinfo {author} {\bibfnamefont {Q.~X.}\ \bibnamefont {Jia}}, \bibinfo {author} {\bibfnamefont {B.}~\bibnamefont {Maiorov}}, \bibinfo {author} {\bibfnamefont {H.}~\bibnamefont {Wang}},\ and\ \bibinfo {author} {\bibfnamefont {M.}~\bibnamefont {Maley}},\ }\bibfield  {title} {\bibinfo {title} {Materials science challenges for high-temperature superconducting wire},\ }\href {https://doi.org/10.1038/nmat1989} {\bibfield  {journal} {\bibinfo  {journal} {Nat. Mater.}\ }\textbf {\bibinfo {volume} {6}},\ \bibinfo {pages} {631} (\bibinfo {year} {2007})}\BibitemShut {NoStop}%
\bibitem [{\citenamefont {Hilgenkamp}\ and\ \citenamefont {Mannhart}(2002)}]{hilgenkamp2002grain}%
  \BibitemOpen
  \bibfield  {author} {\bibinfo {author} {\bibfnamefont {H.}~\bibnamefont {Hilgenkamp}}\ and\ \bibinfo {author} {\bibfnamefont {J.}~\bibnamefont {Mannhart}},\ }\bibfield  {title} {\bibinfo {title} {Grain boundaries in high-${T}_{c}$ superconductors},\ }\href {https://doi.org/10.1103/RevModPhys.74.485} {\bibfield  {journal} {\bibinfo  {journal} {Rev. Mod. Phys.}\ }\textbf {\bibinfo {volume} {74}},\ \bibinfo {pages} {485} (\bibinfo {year} {2002})}\BibitemShut {NoStop}%
\bibitem [{\citenamefont {Graser}\ \emph {et~al.}(2010)\citenamefont {Graser}, \citenamefont {Hirschfeld}, \citenamefont {Kopp}, \citenamefont {Gutser}, \citenamefont {Andersen},\ and\ \citenamefont {Mannhart}}]{graser2010grain}%
  \BibitemOpen
  \bibfield  {author} {\bibinfo {author} {\bibfnamefont {S.}~\bibnamefont {Graser}}, \bibinfo {author} {\bibfnamefont {P.~J.}\ \bibnamefont {Hirschfeld}}, \bibinfo {author} {\bibfnamefont {T.}~\bibnamefont {Kopp}}, \bibinfo {author} {\bibfnamefont {R.}~\bibnamefont {Gutser}}, \bibinfo {author} {\bibfnamefont {B.~M.}\ \bibnamefont {Andersen}},\ and\ \bibinfo {author} {\bibfnamefont {J.}~\bibnamefont {Mannhart}},\ }\bibfield  {title} {\bibinfo {title} {How grain boundaries limit supercurrents in high-temperature superconductors},\ }\href {https://doi.org/10.1038/nphys1687} {\bibfield  {journal} {\bibinfo  {journal} {Nat. Phys.}\ }\textbf {\bibinfo {volume} {6}},\ \bibinfo {pages} {609} (\bibinfo {year} {2010})}\BibitemShut {NoStop}%
\bibitem [{\citenamefont {Reade}\ \emph {et~al.}(1992)\citenamefont {Reade}, \citenamefont {Berdahl}, \citenamefont {Russo},\ and\ \citenamefont {Garrison}}]{reade1992laser}%
  \BibitemOpen
  \bibfield  {author} {\bibinfo {author} {\bibfnamefont {R.~P.}\ \bibnamefont {Reade}}, \bibinfo {author} {\bibfnamefont {P.}~\bibnamefont {Berdahl}}, \bibinfo {author} {\bibfnamefont {R.~E.}\ \bibnamefont {Russo}},\ and\ \bibinfo {author} {\bibfnamefont {S.~M.}\ \bibnamefont {Garrison}},\ }\bibfield  {title} {\bibinfo {title} {Laser deposition of biaxially textured yttria‐stabilized zirconia buffer layers on polycrystalline metallic alloys for high critical current {Y‐Ba‐Cu‐O} thin films},\ }\href {https://doi.org/10.1063/1.108277} {\bibfield  {journal} {\bibinfo  {journal} {Appl. Phys. Lett.}\ }\textbf {\bibinfo {volume} {61}},\ \bibinfo {pages} {2231} (\bibinfo {year} {1992})}\BibitemShut {NoStop}%
\bibitem [{\citenamefont {Iijima}\ \emph {et~al.}(1993)\citenamefont {Iijima}, \citenamefont {Onabe}, \citenamefont {Futaki}, \citenamefont {Tanabe}, \citenamefont {Sadakata}, \citenamefont {Kohno},\ and\ \citenamefont {Ikeno}}]{ijima1993structural}%
  \BibitemOpen
  \bibfield  {author} {\bibinfo {author} {\bibfnamefont {Y.}~\bibnamefont {Iijima}}, \bibinfo {author} {\bibfnamefont {K.}~\bibnamefont {Onabe}}, \bibinfo {author} {\bibfnamefont {N.}~\bibnamefont {Futaki}}, \bibinfo {author} {\bibfnamefont {N.}~\bibnamefont {Tanabe}}, \bibinfo {author} {\bibfnamefont {N.}~\bibnamefont {Sadakata}}, \bibinfo {author} {\bibfnamefont {O.}~\bibnamefont {Kohno}},\ and\ \bibinfo {author} {\bibfnamefont {Y.}~\bibnamefont {Ikeno}},\ }\bibfield  {title} {\bibinfo {title} {Structural and transport properties of biaxially aligned {YBa$_2$Cu$_3$O$_{7-x}$} films on polycrystalline {Ni}‐based alloy with ion‐beam‐modified buffer layers},\ }\href {https://doi.org/10.1063/1.354801} {\bibfield  {journal} {\bibinfo  {journal} {J. Appl. Phys.}\ }\textbf {\bibinfo {volume} {74}},\ \bibinfo {pages} {1905} (\bibinfo {year} {1993})}\BibitemShut {NoStop}%
\bibitem [{\citenamefont {Wu}\ \emph {et~al.}(1995)\citenamefont {Wu}, \citenamefont {Foltyn}, \citenamefont {Arendt}, \citenamefont {Blumenthal}, \citenamefont {Campbell}, \citenamefont {Cotton}, \citenamefont {Coulter}, \citenamefont {Hults}, \citenamefont {Maley}, \citenamefont {Safar},\ and\ \citenamefont {Smith}}]{wu1995properties}%
  \BibitemOpen
  \bibfield  {author} {\bibinfo {author} {\bibfnamefont {X.~D.}\ \bibnamefont {Wu}}, \bibinfo {author} {\bibfnamefont {S.~R.}\ \bibnamefont {Foltyn}}, \bibinfo {author} {\bibfnamefont {P.~N.}\ \bibnamefont {Arendt}}, \bibinfo {author} {\bibfnamefont {W.~R.}\ \bibnamefont {Blumenthal}}, \bibinfo {author} {\bibfnamefont {I.~H.}\ \bibnamefont {Campbell}}, \bibinfo {author} {\bibfnamefont {J.~D.}\ \bibnamefont {Cotton}}, \bibinfo {author} {\bibfnamefont {J.~Y.}\ \bibnamefont {Coulter}}, \bibinfo {author} {\bibfnamefont {W.~L.}\ \bibnamefont {Hults}}, \bibinfo {author} {\bibfnamefont {M.~P.}\ \bibnamefont {Maley}}, \bibinfo {author} {\bibfnamefont {H.~F.}\ \bibnamefont {Safar}},\ and\ \bibinfo {author} {\bibfnamefont {J.~L.}\ \bibnamefont {Smith}},\ }\bibfield  {title} {\bibinfo {title} {Properties of {YBa$_2$Cu$_3$O$_{7-\delta}$} thick films on flexible buffered metallic substrates},\ }\href {https://doi.org/10.1063/1.114559} {\bibfield  {journal} {\bibinfo  {journal} {Appl. Phys. Lett.}\ }\textbf {\bibinfo
  {volume} {67}},\ \bibinfo {pages} {2397} (\bibinfo {year} {1995})}\BibitemShut {NoStop}%
\bibitem [{\citenamefont {Goyal}\ \emph {et~al.}(1996)\citenamefont {Goyal}, \citenamefont {Norton}, \citenamefont {Budai}, \citenamefont {Paranthaman}, \citenamefont {Specht}, \citenamefont {Kroeger}, \citenamefont {Christen}, \citenamefont {He}, \citenamefont {Saffian}, \citenamefont {List}, \citenamefont {Lee}, \citenamefont {Martin}, \citenamefont {Klabunde}, \citenamefont {Hartfield},\ and\ \citenamefont {Sikka}}]{goyal1996high}%
  \BibitemOpen
  \bibfield  {author} {\bibinfo {author} {\bibfnamefont {A.}~\bibnamefont {Goyal}}, \bibinfo {author} {\bibfnamefont {D.~P.}\ \bibnamefont {Norton}}, \bibinfo {author} {\bibfnamefont {J.~D.}\ \bibnamefont {Budai}}, \bibinfo {author} {\bibfnamefont {M.}~\bibnamefont {Paranthaman}}, \bibinfo {author} {\bibfnamefont {E.~D.}\ \bibnamefont {Specht}}, \bibinfo {author} {\bibfnamefont {D.~M.}\ \bibnamefont {Kroeger}}, \bibinfo {author} {\bibfnamefont {D.~K.}\ \bibnamefont {Christen}}, \bibinfo {author} {\bibfnamefont {Q.}~\bibnamefont {He}}, \bibinfo {author} {\bibfnamefont {B.}~\bibnamefont {Saffian}}, \bibinfo {author} {\bibfnamefont {F.~A.}\ \bibnamefont {List}}, \bibinfo {author} {\bibfnamefont {D.~F.}\ \bibnamefont {Lee}}, \bibinfo {author} {\bibfnamefont {P.~M.}\ \bibnamefont {Martin}}, \bibinfo {author} {\bibfnamefont {C.~E.}\ \bibnamefont {Klabunde}}, \bibinfo {author} {\bibfnamefont {E.}~\bibnamefont {Hartfield}},\ and\ \bibinfo {author} {\bibfnamefont {V.~K.}\ \bibnamefont {Sikka}},\ }\bibfield  {title}
  {\bibinfo {title} {High critical current density superconducting tapes by epitaxial deposition of {YBa$_2$Cu$_3$O$_{x}$} thick films on biaxially textured metals},\ }\href {https://doi.org/10.1063/1.117489} {\bibfield  {journal} {\bibinfo  {journal} {Appl. Phys. Lett.}\ }\textbf {\bibinfo {volume} {69}},\ \bibinfo {pages} {1795} (\bibinfo {year} {1996})}\BibitemShut {NoStop}%
\bibitem [{\citenamefont {D{\'\i}ez-Sierra}\ \emph {et~al.}(2020)\citenamefont {D{\'\i}ez-Sierra} \emph {et~al.}}]{diez2020high}%
  \BibitemOpen
  \bibfield  {author} {\bibinfo {author} {\bibfnamefont {J.}~\bibnamefont {D{\'\i}ez-Sierra}} \emph {et~al.},\ }\bibfield  {title} {\bibinfo {title} {High critical current density and enhanced pinning in superconducting films of {YBa$_2$Cu$_3$O$_{7-\delta}$} nanocomposites with embedded {BaZrO$_3$}, {BaHfO$_3$}, {BaTiO$_3$}, and {SrZrO$_3$} nanocrystals},\ }\href {https://doi.org/10.1021/acsanm.0c00814} {\bibfield  {journal} {\bibinfo  {journal} {ACS Appl. Nano Mater.}\ }\textbf {\bibinfo {volume} {3}},\ \bibinfo {pages} {5542} (\bibinfo {year} {2020})}\BibitemShut {NoStop}%
\bibitem [{\citenamefont {Ahmad}\ \emph {et~al.}(2021)\citenamefont {Ahmad}, \citenamefont {Sarangi},\ and\ \citenamefont {Sarun}}]{ahmad2021enhanced}%
  \BibitemOpen
  \bibfield  {author} {\bibinfo {author} {\bibfnamefont {I.}~\bibnamefont {Ahmad}}, \bibinfo {author} {\bibfnamefont {S.}~\bibnamefont {Sarangi}},\ and\ \bibinfo {author} {\bibfnamefont {P.}~\bibnamefont {Sarun}},\ }\bibfield  {title} {\bibinfo {title} {Enhanced critical current density and flux pinning of anthracene doped magnesium diboride superconductor},\ }\href {https://doi.org/https://doi.org/10.1016/j.jallcom.2021.160999} {\bibfield  {journal} {\bibinfo  {journal} {J. Alloys Compd.}\ }\textbf {\bibinfo {volume} {884}},\ \bibinfo {pages} {160999} (\bibinfo {year} {2021})}\BibitemShut {NoStop}%
\bibitem [{\citenamefont {Rocci}\ \emph {et~al.}(2020)\citenamefont {Rocci}, \citenamefont {Suri}, \citenamefont {Kamra}, \citenamefont {Vilela}, \citenamefont {Takamura}, \citenamefont {Nemes}, \citenamefont {Martinez}, \citenamefont {Hernandez},\ and\ \citenamefont {Moodera}}]{rocci2020large}%
  \BibitemOpen
  \bibfield  {author} {\bibinfo {author} {\bibfnamefont {M.}~\bibnamefont {Rocci}}, \bibinfo {author} {\bibfnamefont {D.}~\bibnamefont {Suri}}, \bibinfo {author} {\bibfnamefont {A.}~\bibnamefont {Kamra}}, \bibinfo {author} {\bibfnamefont {G.}~\bibnamefont {Vilela}}, \bibinfo {author} {\bibfnamefont {Y.}~\bibnamefont {Takamura}}, \bibinfo {author} {\bibfnamefont {N.~M.}\ \bibnamefont {Nemes}}, \bibinfo {author} {\bibfnamefont {J.~L.}\ \bibnamefont {Martinez}}, \bibinfo {author} {\bibfnamefont {M.~G.}\ \bibnamefont {Hernandez}},\ and\ \bibinfo {author} {\bibfnamefont {J.~S.}\ \bibnamefont {Moodera}},\ }\bibfield  {title} {\bibinfo {title} {Large enhancement of critical current in superconducting devices by gate voltage},\ }\href {https://doi.org/10.1021/acs.nanolett.0c03547} {\bibfield  {journal} {\bibinfo  {journal} {Nano Lett.}\ }\textbf {\bibinfo {volume} {21}},\ \bibinfo {pages} {216} (\bibinfo {year} {2020})}\BibitemShut {NoStop}%
\bibitem [{\citenamefont {Caffer}\ \emph {et~al.}(2021)\citenamefont {Caffer}, \citenamefont {Chaves}, \citenamefont {Pessoa}, \citenamefont {Carvalho}, \citenamefont {Ortiz}, \citenamefont {Zadorosny},\ and\ \citenamefont {Motta}}]{caffer2021optimum}%
  \BibitemOpen
  \bibfield  {author} {\bibinfo {author} {\bibfnamefont {A.~M.}\ \bibnamefont {Caffer}}, \bibinfo {author} {\bibfnamefont {D.~A.~D.}\ \bibnamefont {Chaves}}, \bibinfo {author} {\bibfnamefont {A.~L.}\ \bibnamefont {Pessoa}}, \bibinfo {author} {\bibfnamefont {C.~L.}\ \bibnamefont {Carvalho}}, \bibinfo {author} {\bibfnamefont {W.~A.}\ \bibnamefont {Ortiz}}, \bibinfo {author} {\bibfnamefont {R.}~\bibnamefont {Zadorosny}},\ and\ \bibinfo {author} {\bibfnamefont {M.}~\bibnamefont {Motta}},\ }\bibfield  {title} {\bibinfo {title} {Optimum heat treatment to enhance the weak-link response of {Y123} nanowires prepared by solution blow spinning},\ }\href {https://doi.org/10.1088/1361-6668/abc8d2} {\bibfield  {journal} {\bibinfo  {journal} {Supercond. Sci. Technol.}\ }\textbf {\bibinfo {volume} {34}},\ \bibinfo {pages} {025009} (\bibinfo {year} {2021})}\BibitemShut {NoStop}%
\bibitem [{\citenamefont {Algarni}\ \emph {et~al.}(2021)\citenamefont {Algarni}, \citenamefont {Almessiere}, \citenamefont {Slimani}, \citenamefont {Hannachi},\ and\ \citenamefont {Azzouz}}]{algarni2021enhanced}%
  \BibitemOpen
  \bibfield  {author} {\bibinfo {author} {\bibfnamefont {R.}~\bibnamefont {Algarni}}, \bibinfo {author} {\bibfnamefont {M.~A.}\ \bibnamefont {Almessiere}}, \bibinfo {author} {\bibfnamefont {Y.}~\bibnamefont {Slimani}}, \bibinfo {author} {\bibfnamefont {E.}~\bibnamefont {Hannachi}},\ and\ \bibinfo {author} {\bibfnamefont {F.~B.}\ \bibnamefont {Azzouz}},\ }\bibfield  {title} {\bibinfo {title} {Enhanced critical current density and flux pinning traits with {Dy$_2$O$_3$} nanoparticles added to {YBa$_2$Cu$_3$O$_{7-d}$} superconductor},\ }\href {https://doi.org/10.1016/j.jallcom.2020.157019} {\bibfield  {journal} {\bibinfo  {journal} {J. Alloys Compd.}\ }\textbf {\bibinfo {volume} {852}},\ \bibinfo {pages} {157019} (\bibinfo {year} {2021})}\BibitemShut {NoStop}%
\bibitem [{\citenamefont {Chaves}\ \emph {et~al.}(2021)\citenamefont {Chaves}, \citenamefont {de~Araújo}, \citenamefont {Carmo}, \citenamefont {Colauto}, \citenamefont {Oliveira}, \citenamefont {Andrade}, \citenamefont {Johansen}, \citenamefont {Silhanek}, \citenamefont {Ortiz},\ and\ \citenamefont {Motta}}]{chaves2021enhancing}%
  \BibitemOpen
  \bibfield  {author} {\bibinfo {author} {\bibfnamefont {D.~A.~D.}\ \bibnamefont {Chaves}}, \bibinfo {author} {\bibfnamefont {I.~M.}\ \bibnamefont {de~Araújo}}, \bibinfo {author} {\bibfnamefont {D.}~\bibnamefont {Carmo}}, \bibinfo {author} {\bibfnamefont {F.}~\bibnamefont {Colauto}}, \bibinfo {author} {\bibfnamefont {A.~A.~M.}\ \bibnamefont {Oliveira}}, \bibinfo {author} {\bibfnamefont {A.~M.~H.}\ \bibnamefont {Andrade}}, \bibinfo {author} {\bibfnamefont {T.~H.}\ \bibnamefont {Johansen}}, \bibinfo {author} {\bibfnamefont {A.~V.}\ \bibnamefont {Silhanek}}, \bibinfo {author} {\bibfnamefont {W.~A.}\ \bibnamefont {Ortiz}},\ and\ \bibinfo {author} {\bibfnamefont {M.}~\bibnamefont {Motta}},\ }\bibfield  {title} {\bibinfo {title} {Enhancing the effective critical current density in a {Nb} superconducting thin film by cooling in an inhomogeneous magnetic field},\ }\href {https://doi.org/10.1063/5.0058680} {\bibfield  {journal} {\bibinfo  {journal} {Appl. Phys. Lett.}\ }\textbf {\bibinfo {volume} {119}},\ \bibinfo
  {pages} {022602} (\bibinfo {year} {2021})}\BibitemShut {NoStop}%
\bibitem [{\citenamefont {Shaw}\ \emph {et~al.}(2018)\citenamefont {Shaw}, \citenamefont {Brisbois}, \citenamefont {Pinheiro}, \citenamefont {Müller}, \citenamefont {Blanco~Alvarez}, \citenamefont {Devillers}, \citenamefont {Dempsey}, \citenamefont {Scheerder}, \citenamefont {Van~de Vondel}, \citenamefont {Melinte}, \citenamefont {Vanderbemden}, \citenamefont {Motta}, \citenamefont {Ortiz}, \citenamefont {Hasselbach}, \citenamefont {Kramer},\ and\ \citenamefont {Silhanek}}]{shaw2018quantitative}%
  \BibitemOpen
  \bibfield  {author} {\bibinfo {author} {\bibfnamefont {G.}~\bibnamefont {Shaw}}, \bibinfo {author} {\bibfnamefont {J.}~\bibnamefont {Brisbois}}, \bibinfo {author} {\bibfnamefont {L.~B. G.~L.}\ \bibnamefont {Pinheiro}}, \bibinfo {author} {\bibfnamefont {J.}~\bibnamefont {Müller}}, \bibinfo {author} {\bibfnamefont {S.}~\bibnamefont {Blanco~Alvarez}}, \bibinfo {author} {\bibfnamefont {T.}~\bibnamefont {Devillers}}, \bibinfo {author} {\bibfnamefont {N.~M.}\ \bibnamefont {Dempsey}}, \bibinfo {author} {\bibfnamefont {J.~E.}\ \bibnamefont {Scheerder}}, \bibinfo {author} {\bibfnamefont {J.}~\bibnamefont {Van~de Vondel}}, \bibinfo {author} {\bibfnamefont {S.}~\bibnamefont {Melinte}}, \bibinfo {author} {\bibfnamefont {P.}~\bibnamefont {Vanderbemden}}, \bibinfo {author} {\bibfnamefont {M.}~\bibnamefont {Motta}}, \bibinfo {author} {\bibfnamefont {W.~A.}\ \bibnamefont {Ortiz}}, \bibinfo {author} {\bibfnamefont {K.}~\bibnamefont {Hasselbach}}, \bibinfo {author} {\bibfnamefont {R.~B.~G.}\ \bibnamefont {Kramer}},\ and\
  \bibinfo {author} {\bibfnamefont {A.~V.}\ \bibnamefont {Silhanek}},\ }\bibfield  {title} {\bibinfo {title} {Quantitative magneto-optical investigation of superconductor/ferromagnet hybrid structures},\ }\href {https://doi.org/10.1063/1.5016293} {\bibfield  {journal} {\bibinfo  {journal} {Rev. Sci. Instrum.}\ }\textbf {\bibinfo {volume} {89}},\ \bibinfo {pages} {023705} (\bibinfo {year} {2018})}\BibitemShut {NoStop}%
\bibitem [{\citenamefont {Meltzer}\ \emph {et~al.}(2017)\citenamefont {Meltzer}, \citenamefont {Levin},\ and\ \citenamefont {Zeldov}}]{meltzer2017direct}%
  \BibitemOpen
  \bibfield  {author} {\bibinfo {author} {\bibfnamefont {A.~Y.}\ \bibnamefont {Meltzer}}, \bibinfo {author} {\bibfnamefont {E.}~\bibnamefont {Levin}},\ and\ \bibinfo {author} {\bibfnamefont {E.}~\bibnamefont {Zeldov}},\ }\bibfield  {title} {\bibinfo {title} {Direct reconstruction of two-dimensional currents in thin films from magnetic-field measurements},\ }\href {https://doi.org/10.1103/PhysRevApplied.8.064030} {\bibfield  {journal} {\bibinfo  {journal} {Phys. Rev. Applied}\ }\textbf {\bibinfo {volume} {8}},\ \bibinfo {pages} {064030} (\bibinfo {year} {2017})}\BibitemShut {NoStop}%
\bibitem [{\citenamefont {Vestg\aa{}rden}\ \emph {et~al.}(2011)\citenamefont {Vestg\aa{}rden}, \citenamefont {Shantsev}, \citenamefont {Galperin},\ and\ \citenamefont {Johansen}}]{vestgarden2011dynamics}%
  \BibitemOpen
  \bibfield  {author} {\bibinfo {author} {\bibfnamefont {J.~I.}\ \bibnamefont {Vestg\aa{}rden}}, \bibinfo {author} {\bibfnamefont {D.~V.}\ \bibnamefont {Shantsev}}, \bibinfo {author} {\bibfnamefont {Y.~M.}\ \bibnamefont {Galperin}},\ and\ \bibinfo {author} {\bibfnamefont {T.~H.}\ \bibnamefont {Johansen}},\ }\bibfield  {title} {\bibinfo {title} {Dynamics and morphology of dendritic flux avalanches in superconducting films},\ }\href {https://doi.org/10.1103/PhysRevB.84.054537} {\bibfield  {journal} {\bibinfo  {journal} {Phys. Rev. B}\ }\textbf {\bibinfo {volume} {84}},\ \bibinfo {pages} {054537} (\bibinfo {year} {2011})}\BibitemShut {NoStop}%
\bibitem [{\citenamefont {Jiang}\ \emph {et~al.}(2020)\citenamefont {Jiang}, \citenamefont {Xue}, \citenamefont {Burger}, \citenamefont {Vanderheyden}, \citenamefont {Silhanek},\ and\ \citenamefont {Zhou}}]{jiang2020selective}%
  \BibitemOpen
  \bibfield  {author} {\bibinfo {author} {\bibfnamefont {L.}~\bibnamefont {Jiang}}, \bibinfo {author} {\bibfnamefont {C.}~\bibnamefont {Xue}}, \bibinfo {author} {\bibfnamefont {L.}~\bibnamefont {Burger}}, \bibinfo {author} {\bibfnamefont {B.}~\bibnamefont {Vanderheyden}}, \bibinfo {author} {\bibfnamefont {A.~V.}\ \bibnamefont {Silhanek}},\ and\ \bibinfo {author} {\bibfnamefont {Y.-H.}\ \bibnamefont {Zhou}},\ }\bibfield  {title} {\bibinfo {title} {Selective triggering of magnetic flux avalanches by an edge indentation},\ }\href {https://doi.org/10.1103/PhysRevB.101.224505} {\bibfield  {journal} {\bibinfo  {journal} {Phys. Rev. B}\ }\textbf {\bibinfo {volume} {101}},\ \bibinfo {pages} {224505} (\bibinfo {year} {2020})}\BibitemShut {NoStop}%
\bibitem [{\citenamefont {Schuster}\ \emph {et~al.}(1994{\natexlab{a}})\citenamefont {Schuster}, \citenamefont {Indenbom}, \citenamefont {Koblischka}, \citenamefont {Kuhn},\ and\ \citenamefont {Kronm\"uller}}]{schuster1994observation}%
  \BibitemOpen
  \bibfield  {author} {\bibinfo {author} {\bibfnamefont {T.}~\bibnamefont {Schuster}}, \bibinfo {author} {\bibfnamefont {M.~V.}\ \bibnamefont {Indenbom}}, \bibinfo {author} {\bibfnamefont {M.~R.}\ \bibnamefont {Koblischka}}, \bibinfo {author} {\bibfnamefont {H.}~\bibnamefont {Kuhn}},\ and\ \bibinfo {author} {\bibfnamefont {H.}~\bibnamefont {Kronm\"uller}},\ }\bibfield  {title} {\bibinfo {title} {Observation of current-discontinuity lines in type-{II} superconductors},\ }\href {https://doi.org/10.1103/PhysRevB.49.3443} {\bibfield  {journal} {\bibinfo  {journal} {Phys. Rev. B}\ }\textbf {\bibinfo {volume} {49}},\ \bibinfo {pages} {3443} (\bibinfo {year} {1994}{\natexlab{a}})}\BibitemShut {NoStop}%
\bibitem [{\citenamefont {Schuster}\ \emph {et~al.}(1995)\citenamefont {Schuster}, \citenamefont {Kuhn},\ and\ \citenamefont {Indenbom}}]{schuster1995discontinuity}%
  \BibitemOpen
  \bibfield  {author} {\bibinfo {author} {\bibfnamefont {T.}~\bibnamefont {Schuster}}, \bibinfo {author} {\bibfnamefont {H.}~\bibnamefont {Kuhn}},\ and\ \bibinfo {author} {\bibfnamefont {M.~V.}\ \bibnamefont {Indenbom}},\ }\bibfield  {title} {\bibinfo {title} {Discontinuity lines in rectangular superconductors with intrinsic and extrinsic anisotropies},\ }\href {https://doi.org/10.1103/PhysRevB.52.15621} {\bibfield  {journal} {\bibinfo  {journal} {Phys. Rev. B}\ }\textbf {\bibinfo {volume} {52}},\ \bibinfo {pages} {15621} (\bibinfo {year} {1995})}\BibitemShut {NoStop}%
\bibitem [{\citenamefont {Mints}\ and\ \citenamefont {Rakhmanov}(1981)}]{Mints1981}%
  \BibitemOpen
  \bibfield  {author} {\bibinfo {author} {\bibfnamefont {R.~G.}\ \bibnamefont {Mints}}\ and\ \bibinfo {author} {\bibfnamefont {A.~L.}\ \bibnamefont {Rakhmanov}},\ }\bibfield  {title} {\bibinfo {title} {Critical state stability in type-{II} superconductors and superconducting-normal-metal composites},\ }\href {https://doi.org/10.1103/RevModPhys.53.551} {\bibfield  {journal} {\bibinfo  {journal} {Rev. Mod. Phys.}\ }\textbf {\bibinfo {volume} {53}},\ \bibinfo {pages} {551} (\bibinfo {year} {1981})}\BibitemShut {NoStop}%
\bibitem [{\citenamefont {Blatter}\ \emph {et~al.}(1994)\citenamefont {Blatter}, \citenamefont {Feigel'man}, \citenamefont {Geshkenbein}, \citenamefont {Larkin},\ and\ \citenamefont {Vinokur}}]{Blatter1994}%
  \BibitemOpen
  \bibfield  {author} {\bibinfo {author} {\bibfnamefont {G.}~\bibnamefont {Blatter}}, \bibinfo {author} {\bibfnamefont {M.~V.}\ \bibnamefont {Feigel'man}}, \bibinfo {author} {\bibfnamefont {V.~B.}\ \bibnamefont {Geshkenbein}}, \bibinfo {author} {\bibfnamefont {A.~I.}\ \bibnamefont {Larkin}},\ and\ \bibinfo {author} {\bibfnamefont {V.~M.}\ \bibnamefont {Vinokur}},\ }\bibfield  {title} {\bibinfo {title} {Vortices in high-temperature superconductors},\ }\href {https://doi.org/10.1103/RevModPhys.66.1125} {\bibfield  {journal} {\bibinfo  {journal} {Rev. Mod. Phys.}\ }\textbf {\bibinfo {volume} {66}},\ \bibinfo {pages} {1125} (\bibinfo {year} {1994})}\BibitemShut {NoStop}%
\bibitem [{\citenamefont {Polyanskii}\ \emph {et~al.}(1996)\citenamefont {Polyanskii}, \citenamefont {Gurevich}, \citenamefont {Pashitski}, \citenamefont {Heinig}, \citenamefont {Redwing}, \citenamefont {Nordman},\ and\ \citenamefont {Larbalestier}}]{polyanskii1996magneto}%
  \BibitemOpen
  \bibfield  {author} {\bibinfo {author} {\bibfnamefont {A.~A.}\ \bibnamefont {Polyanskii}}, \bibinfo {author} {\bibfnamefont {A.}~\bibnamefont {Gurevich}}, \bibinfo {author} {\bibfnamefont {A.~E.}\ \bibnamefont {Pashitski}}, \bibinfo {author} {\bibfnamefont {N.~F.}\ \bibnamefont {Heinig}}, \bibinfo {author} {\bibfnamefont {R.~D.}\ \bibnamefont {Redwing}}, \bibinfo {author} {\bibfnamefont {J.~E.}\ \bibnamefont {Nordman}},\ and\ \bibinfo {author} {\bibfnamefont {D.~C.}\ \bibnamefont {Larbalestier}},\ }\bibfield  {title} {\bibinfo {title} {Magneto-optical study of flux penetration and critical current densities in [001] tilt {YBa$_2$Cu$_3$O$_{7-\delta}$} thin-film bicrystals},\ }\href {https://doi.org/10.1103/PhysRevB.53.8687} {\bibfield  {journal} {\bibinfo  {journal} {Phys. Rev. B}\ }\textbf {\bibinfo {volume} {53}},\ \bibinfo {pages} {8687} (\bibinfo {year} {1996})}\BibitemShut {NoStop}%
\bibitem [{\citenamefont {Johansen}\ \emph {et~al.}(2019)\citenamefont {Johansen}, \citenamefont {Colauto}, \citenamefont {de~Andrade}, \citenamefont {Oliveira},\ and\ \citenamefont {Ortiz}}]{johansen2019transparency}%
  \BibitemOpen
  \bibfield  {author} {\bibinfo {author} {\bibfnamefont {T.~H.}\ \bibnamefont {Johansen}}, \bibinfo {author} {\bibfnamefont {F.}~\bibnamefont {Colauto}}, \bibinfo {author} {\bibfnamefont {A.~M.~H.}\ \bibnamefont {de~Andrade}}, \bibinfo {author} {\bibfnamefont {A.~A.~M.}\ \bibnamefont {Oliveira}},\ and\ \bibinfo {author} {\bibfnamefont {W.~A.}\ \bibnamefont {Ortiz}},\ }\bibfield  {title} {\bibinfo {title} {Transparency of planar interfaces in superconductors: a critical-state analysis},\ }\href {https://doi.org/10.1109/TASC.2019.2899209} {\bibfield  {journal} {\bibinfo  {journal} {IEEE Trans. Appl. Supercond.}\ }\textbf {\bibinfo {volume} {29}},\ \bibinfo {pages} {1} (\bibinfo {year} {2019})}\BibitemShut {NoStop}%
\bibitem [{\citenamefont {Colauto}\ \emph {et~al.}(2021)\citenamefont {Colauto}, \citenamefont {Carmo}, \citenamefont {Andrade}, \citenamefont {Oliveira}, \citenamefont {Ortiz}, \citenamefont {Galperin},\ and\ \citenamefont {Johansen}}]{colauto2021measurement}%
  \BibitemOpen
  \bibfield  {author} {\bibinfo {author} {\bibfnamefont {F.}~\bibnamefont {Colauto}}, \bibinfo {author} {\bibfnamefont {D.}~\bibnamefont {Carmo}}, \bibinfo {author} {\bibfnamefont {A.~M.~H.}\ \bibnamefont {Andrade}}, \bibinfo {author} {\bibfnamefont {A.~A.~M.}\ \bibnamefont {Oliveira}}, \bibinfo {author} {\bibfnamefont {W.~A.}\ \bibnamefont {Ortiz}}, \bibinfo {author} {\bibfnamefont {Y.~M.}\ \bibnamefont {Galperin}},\ and\ \bibinfo {author} {\bibfnamefont {T.~H.}\ \bibnamefont {Johansen}},\ }\bibfield  {title} {\bibinfo {title} {Measurement of critical current flow and connectivity in systems of joined square superconducting plates},\ }\href {https://doi.org/10.1016/j.physc.2021.1353931} {\bibfield  {journal} {\bibinfo  {journal} {Physica C}\ }\textbf {\bibinfo {volume} {589}},\ \bibinfo {pages} {1353931} (\bibinfo {year} {2021})}\BibitemShut {NoStop}%
\bibitem [{\citenamefont {Kim}\ \emph {et~al.}(1963)\citenamefont {Kim}, \citenamefont {Hempstead},\ and\ \citenamefont {Strnad}}]{kim1963magnetization}%
  \BibitemOpen
  \bibfield  {author} {\bibinfo {author} {\bibfnamefont {Y.~B.}\ \bibnamefont {Kim}}, \bibinfo {author} {\bibfnamefont {C.~F.}\ \bibnamefont {Hempstead}},\ and\ \bibinfo {author} {\bibfnamefont {A.~R.}\ \bibnamefont {Strnad}},\ }\bibfield  {title} {\bibinfo {title} {Magnetization and critical supercurrents},\ }\href {https://doi.org/10.1103/PhysRev.129.528} {\bibfield  {journal} {\bibinfo  {journal} {Phys. Rev.}\ }\textbf {\bibinfo {volume} {129}},\ \bibinfo {pages} {528} (\bibinfo {year} {1963})}\BibitemShut {NoStop}%
\bibitem [{\citenamefont {Shantsev}\ \emph {et~al.}(2000)\citenamefont {Shantsev}, \citenamefont {Galperin},\ and\ \citenamefont {Johansen}}]{shantsev2000thin}%
  \BibitemOpen
  \bibfield  {author} {\bibinfo {author} {\bibfnamefont {D.~V.}\ \bibnamefont {Shantsev}}, \bibinfo {author} {\bibfnamefont {Y.~M.}\ \bibnamefont {Galperin}},\ and\ \bibinfo {author} {\bibfnamefont {T.~H.}\ \bibnamefont {Johansen}},\ }\bibfield  {title} {\bibinfo {title} {Thin superconducting disk with field-dependent critical current: Magnetization and ac susceptibilities},\ }\href {https://doi.org/10.1103/PhysRevB.61.9699} {\bibfield  {journal} {\bibinfo  {journal} {Phys. Rev. B}\ }\textbf {\bibinfo {volume} {61}},\ \bibinfo {pages} {9699} (\bibinfo {year} {2000})}\BibitemShut {NoStop}%
\bibitem [{\citenamefont {Johansen}\ and\ \citenamefont {Bratsberg}(1995)}]{johansen1995critical}%
  \BibitemOpen
  \bibfield  {author} {\bibinfo {author} {\bibfnamefont {T.~H.}\ \bibnamefont {Johansen}}\ and\ \bibinfo {author} {\bibfnamefont {H.}~\bibnamefont {Bratsberg}},\ }\bibfield  {title} {\bibinfo {title} {Critical‐state magnetization of type‐{II} superconductors in rectangular slab and cylinder geometries},\ }\href {https://doi.org/10.1063/1.358576} {\bibfield  {journal} {\bibinfo  {journal} {J. Appl. Phys.}\ }\textbf {\bibinfo {volume} {77}},\ \bibinfo {pages} {3945} (\bibinfo {year} {1995})}\BibitemShut {NoStop}%
\bibitem [{\citenamefont {Brandt}\ and\ \citenamefont {Indenbom}(1993)}]{brandt1993Type}%
  \BibitemOpen
  \bibfield  {author} {\bibinfo {author} {\bibfnamefont {E.~H.}\ \bibnamefont {Brandt}}\ and\ \bibinfo {author} {\bibfnamefont {M.}~\bibnamefont {Indenbom}},\ }\bibfield  {title} {\bibinfo {title} {Type-{II}-superconductor strip with current in a perpendicular magnetic field},\ }\href {https://doi.org/10.1103/PhysRevB.48.12893} {\bibfield  {journal} {\bibinfo  {journal} {Phys. Rev. B}\ }\textbf {\bibinfo {volume} {48}},\ \bibinfo {pages} {12893} (\bibinfo {year} {1993})}\BibitemShut {NoStop}%
\bibitem [{\citenamefont {Shantsev}\ \emph {et~al.}(1999)\citenamefont {Shantsev}, \citenamefont {Koblischka}, \citenamefont {Galperin}, \citenamefont {Johansen}, \citenamefont {P\ifmmode~\mathring{u}\else \r{u}\fi{}st},\ and\ \citenamefont {Jirsa}}]{shantsev1999central}%
  \BibitemOpen
  \bibfield  {author} {\bibinfo {author} {\bibfnamefont {D.~V.}\ \bibnamefont {Shantsev}}, \bibinfo {author} {\bibfnamefont {M.~R.}\ \bibnamefont {Koblischka}}, \bibinfo {author} {\bibfnamefont {Y.~M.}\ \bibnamefont {Galperin}}, \bibinfo {author} {\bibfnamefont {T.~H.}\ \bibnamefont {Johansen}}, \bibinfo {author} {\bibfnamefont {L.}~\bibnamefont {P\ifmmode~\mathring{u}\else \r{u}\fi{}st}},\ and\ \bibinfo {author} {\bibfnamefont {M.}~\bibnamefont {Jirsa}},\ }\bibfield  {title} {\bibinfo {title} {Central peak position in magnetization loops of high- ${T}_{c}$ superconductors},\ }\href {https://doi.org/10.1103/PhysRevLett.82.2947} {\bibfield  {journal} {\bibinfo  {journal} {Phys. Rev. Lett.}\ }\textbf {\bibinfo {volume} {82}},\ \bibinfo {pages} {2947} (\bibinfo {year} {1999})}\BibitemShut {NoStop}%
\bibitem [{\citenamefont {Wambaugh}\ \emph {et~al.}(1999)\citenamefont {Wambaugh}, \citenamefont {Reichhardt}, \citenamefont {Olson}, \citenamefont {Marchesoni},\ and\ \citenamefont {Nori}}]{Wambaugh1999}%
  \BibitemOpen
  \bibfield  {author} {\bibinfo {author} {\bibfnamefont {J.~F.}\ \bibnamefont {Wambaugh}}, \bibinfo {author} {\bibfnamefont {C.}~\bibnamefont {Reichhardt}}, \bibinfo {author} {\bibfnamefont {C.~J.}\ \bibnamefont {Olson}}, \bibinfo {author} {\bibfnamefont {F.}~\bibnamefont {Marchesoni}},\ and\ \bibinfo {author} {\bibfnamefont {F.}~\bibnamefont {Nori}},\ }\bibfield  {title} {\bibinfo {title} {Superconducting fluxon pumps and lenses},\ }\href {https://doi.org/10.1103/PhysRevLett.83.5106} {\bibfield  {journal} {\bibinfo  {journal} {Phys. Rev. Lett.}\ }\textbf {\bibinfo {volume} {83}},\ \bibinfo {pages} {5106} (\bibinfo {year} {1999})}\BibitemShut {NoStop}%
\bibitem [{\citenamefont {Olson}\ \emph {et~al.}(2001)\citenamefont {Olson}, \citenamefont {Reichhardt}, \citenamefont {Jank\'o},\ and\ \citenamefont {Nori}}]{Olson2001}%
  \BibitemOpen
  \bibfield  {author} {\bibinfo {author} {\bibfnamefont {C.~J.}\ \bibnamefont {Olson}}, \bibinfo {author} {\bibfnamefont {C.}~\bibnamefont {Reichhardt}}, \bibinfo {author} {\bibfnamefont {B.}~\bibnamefont {Jank\'o}},\ and\ \bibinfo {author} {\bibfnamefont {F.}~\bibnamefont {Nori}},\ }\bibfield  {title} {\bibinfo {title} {Collective interaction-driven ratchet for transporting flux quanta},\ }\href {https://doi.org/10.1103/PhysRevLett.87.177002} {\bibfield  {journal} {\bibinfo  {journal} {Phys. Rev. Lett.}\ }\textbf {\bibinfo {volume} {87}},\ \bibinfo {pages} {177002} (\bibinfo {year} {2001})}\BibitemShut {NoStop}%
\bibitem [{\citenamefont {Togawa}\ \emph {et~al.}(2005)\citenamefont {Togawa}, \citenamefont {Harada}, \citenamefont {Akashi}, \citenamefont {Kasai}, \citenamefont {Matsuda}, \citenamefont {Nori}, \citenamefont {Maeda},\ and\ \citenamefont {Tonomura}}]{Togawa2005}%
  \BibitemOpen
  \bibfield  {author} {\bibinfo {author} {\bibfnamefont {Y.}~\bibnamefont {Togawa}}, \bibinfo {author} {\bibfnamefont {K.}~\bibnamefont {Harada}}, \bibinfo {author} {\bibfnamefont {T.}~\bibnamefont {Akashi}}, \bibinfo {author} {\bibfnamefont {H.}~\bibnamefont {Kasai}}, \bibinfo {author} {\bibfnamefont {T.}~\bibnamefont {Matsuda}}, \bibinfo {author} {\bibfnamefont {F.}~\bibnamefont {Nori}}, \bibinfo {author} {\bibfnamefont {A.}~\bibnamefont {Maeda}},\ and\ \bibinfo {author} {\bibfnamefont {A.}~\bibnamefont {Tonomura}},\ }\bibfield  {title} {\bibinfo {title} {Direct observation of rectified motion of vortices in a niobium superconductor},\ }\href {https://doi.org/10.1103/PhysRevLett.95.087002} {\bibfield  {journal} {\bibinfo  {journal} {Phys. Rev. Lett.}\ }\textbf {\bibinfo {volume} {95}},\ \bibinfo {pages} {087002} (\bibinfo {year} {2005})}\BibitemShut {NoStop}%
\bibitem [{\citenamefont {de~Souza~Silva}\ \emph {et~al.}(2006{\natexlab{a}})\citenamefont {de~Souza~Silva}, \citenamefont {Van~de Vondel}, \citenamefont {Zhu}, \citenamefont {Morelle},\ and\ \citenamefont {Moshchalkov}}]{Clecio2006a}%
  \BibitemOpen
  \bibfield  {author} {\bibinfo {author} {\bibfnamefont {C.~C.}\ \bibnamefont {de~Souza~Silva}}, \bibinfo {author} {\bibfnamefont {J.}~\bibnamefont {Van~de Vondel}}, \bibinfo {author} {\bibfnamefont {B.~Y.}\ \bibnamefont {Zhu}}, \bibinfo {author} {\bibfnamefont {M.}~\bibnamefont {Morelle}},\ and\ \bibinfo {author} {\bibfnamefont {V.~V.}\ \bibnamefont {Moshchalkov}},\ }\bibfield  {title} {\bibinfo {title} {Vortex ratchet effects in films with a periodic array of antidots},\ }\href {https://doi.org/10.1103/PhysRevB.73.014507} {\bibfield  {journal} {\bibinfo  {journal} {Phys. Rev. B}\ }\textbf {\bibinfo {volume} {73}},\ \bibinfo {pages} {014507} (\bibinfo {year} {2006}{\natexlab{a}})}\BibitemShut {NoStop}%
\bibitem [{\citenamefont {de~Souza~Silva}\ \emph {et~al.}(2006{\natexlab{b}})\citenamefont {de~Souza~Silva}, \citenamefont {Van~de Vondel}, \citenamefont {Morelle},\ and\ \citenamefont {Moshchalkov}}]{Clecio2006b}%
  \BibitemOpen
  \bibfield  {author} {\bibinfo {author} {\bibfnamefont {C.~C.}\ \bibnamefont {de~Souza~Silva}}, \bibinfo {author} {\bibfnamefont {J.}~\bibnamefont {Van~de Vondel}}, \bibinfo {author} {\bibfnamefont {M.}~\bibnamefont {Morelle}},\ and\ \bibinfo {author} {\bibfnamefont {V.~V.}\ \bibnamefont {Moshchalkov}},\ }\bibfield  {title} {\bibinfo {title} {Controlled multiple reversals of a ratchet effect},\ }\href {https://doi.org/10.1038/nature04595} {\bibfield  {journal} {\bibinfo  {journal} {Nature}\ }\textbf {\bibinfo {volume} {440}},\ \bibinfo {pages} {651} (\bibinfo {year} {2006}{\natexlab{b}})}\BibitemShut {NoStop}%
\bibitem [{\citenamefont {Schuster}\ \emph {et~al.}(1994{\natexlab{b}})\citenamefont {Schuster}, \citenamefont {Indenbom}, \citenamefont {Koblischka}, \citenamefont {Kuhn},\ and\ \citenamefont {Kronm\"uller}}]{Schuster1994}%
  \BibitemOpen
  \bibfield  {author} {\bibinfo {author} {\bibfnamefont {T.}~\bibnamefont {Schuster}}, \bibinfo {author} {\bibfnamefont {M.~V.}\ \bibnamefont {Indenbom}}, \bibinfo {author} {\bibfnamefont {M.~R.}\ \bibnamefont {Koblischka}}, \bibinfo {author} {\bibfnamefont {H.}~\bibnamefont {Kuhn}},\ and\ \bibinfo {author} {\bibfnamefont {H.}~\bibnamefont {Kronm\"uller}},\ }\bibfield  {title} {\bibinfo {title} {Observation of current-discontinuity lines in type-{II} superconductors},\ }\href {https://doi.org/10.1103/PhysRevB.49.3443} {\bibfield  {journal} {\bibinfo  {journal} {Phys. Rev. B}\ }\textbf {\bibinfo {volume} {49}},\ \bibinfo {pages} {3443} (\bibinfo {year} {1994}{\natexlab{b}})}\BibitemShut {NoStop}%
\bibitem [{\citenamefont {Evetts}\ and\ \citenamefont {Glowacki}(1988)}]{evetts1988relation}%
  \BibitemOpen
  \bibfield  {author} {\bibinfo {author} {\bibfnamefont {J.~E.}\ \bibnamefont {Evetts}}\ and\ \bibinfo {author} {\bibfnamefont {B.~A.}\ \bibnamefont {Glowacki}},\ }\bibfield  {title} {\bibinfo {title} {Relation of critical current irreversibility to trapped flux and microstructure in polycrystalline {YBa$_2$Cu$_3$O$_7$}},\ }\href {https://doi.org/10.1016/0011-2275(88)90147-6} {\bibfield  {journal} {\bibinfo  {journal} {Cryogenics}\ }\textbf {\bibinfo {volume} {28}},\ \bibinfo {pages} {641} (\bibinfo {year} {1988})}\BibitemShut {NoStop}%
\bibitem [{\citenamefont {Dimos}\ \emph {et~al.}(1990)\citenamefont {Dimos}, \citenamefont {Chaudhari},\ and\ \citenamefont {Mannhart}}]{dimos1990superconducting}%
  \BibitemOpen
  \bibfield  {author} {\bibinfo {author} {\bibfnamefont {D.}~\bibnamefont {Dimos}}, \bibinfo {author} {\bibfnamefont {P.}~\bibnamefont {Chaudhari}},\ and\ \bibinfo {author} {\bibfnamefont {J.}~\bibnamefont {Mannhart}},\ }\bibfield  {title} {\bibinfo {title} {Superconducting transport properties of grain boundaries in {YBa$_2$Cu$_3$O$_7$} bicrystals},\ }\href {https://doi.org/10.1103/PhysRevB.41.4038} {\bibfield  {journal} {\bibinfo  {journal} {Phys. Rev. B}\ }\textbf {\bibinfo {volume} {41}},\ \bibinfo {pages} {4038} (\bibinfo {year} {1990})}\BibitemShut {NoStop}%
\bibitem [{\citenamefont {Hogg}\ \emph {et~al.}(2001)\citenamefont {Hogg}, \citenamefont {Kahlmann}, \citenamefont {Barber},\ and\ \citenamefont {Evetts}}]{hogg2001angular}%
  \BibitemOpen
  \bibfield  {author} {\bibinfo {author} {\bibfnamefont {M.~J.}\ \bibnamefont {Hogg}}, \bibinfo {author} {\bibfnamefont {F.}~\bibnamefont {Kahlmann}}, \bibinfo {author} {\bibfnamefont {Z.~H.}\ \bibnamefont {Barber}},\ and\ \bibinfo {author} {\bibfnamefont {J.~E.}\ \bibnamefont {Evetts}},\ }\bibfield  {title} {\bibinfo {title} {Angular hysteresis in the critical current of {YBa$_2$Cu$_3$O$_{7}$} low-angle grain boundaries},\ }\href {https://doi.org/10.1088/0953-2048/14/9/301} {\bibfield  {journal} {\bibinfo  {journal} {Supercond. Sci. Technol.}\ }\textbf {\bibinfo {volume} {14}},\ \bibinfo {pages} {647} (\bibinfo {year} {2001})}\BibitemShut {NoStop}%
\bibitem [{\citenamefont {Thompson}\ \emph {et~al.}(2004)\citenamefont {Thompson}, \citenamefont {Kim}, \citenamefont {Cantoni}, \citenamefont {Christen}, \citenamefont {Feenstra},\ and\ \citenamefont {Verebelyi}}]{thompson2004self}%
  \BibitemOpen
  \bibfield  {author} {\bibinfo {author} {\bibfnamefont {J.~R.}\ \bibnamefont {Thompson}}, \bibinfo {author} {\bibfnamefont {H.~J.}\ \bibnamefont {Kim}}, \bibinfo {author} {\bibfnamefont {C.}~\bibnamefont {Cantoni}}, \bibinfo {author} {\bibfnamefont {D.~K.}\ \bibnamefont {Christen}}, \bibinfo {author} {\bibfnamefont {R.}~\bibnamefont {Feenstra}},\ and\ \bibinfo {author} {\bibfnamefont {D.~T.}\ \bibnamefont {Verebelyi}},\ }\bibfield  {title} {\bibinfo {title} {Self-organized current transport through low-angle grain boundaries in {YBa$_2$Cu$_3$O$_{7-\delta}$} thin films studied magnetometrically},\ }\href {https://doi.org/10.1103/PhysRevB.69.104509} {\bibfield  {journal} {\bibinfo  {journal} {Phys. Rev. B}\ }\textbf {\bibinfo {volume} {69}},\ \bibinfo {pages} {104509} (\bibinfo {year} {2004})}\BibitemShut {NoStop}%
\bibitem [{\citenamefont {Palau}\ \emph {et~al.}(2004)\citenamefont {Palau}, \citenamefont {Puig}, \citenamefont {Obradors}, \citenamefont {Pardo}, \citenamefont {Navau}, \citenamefont {Sanchez}, \citenamefont {Usoskin}, \citenamefont {Freyhardt}, \citenamefont {Fernández}, \citenamefont {Holzapfel},\ and\ \citenamefont {Feenstra}}]{palau2004simultaneous}%
  \BibitemOpen
  \bibfield  {author} {\bibinfo {author} {\bibfnamefont {A.}~\bibnamefont {Palau}}, \bibinfo {author} {\bibfnamefont {T.}~\bibnamefont {Puig}}, \bibinfo {author} {\bibfnamefont {X.}~\bibnamefont {Obradors}}, \bibinfo {author} {\bibfnamefont {E.}~\bibnamefont {Pardo}}, \bibinfo {author} {\bibfnamefont {C.}~\bibnamefont {Navau}}, \bibinfo {author} {\bibfnamefont {A.}~\bibnamefont {Sanchez}}, \bibinfo {author} {\bibfnamefont {A.}~\bibnamefont {Usoskin}}, \bibinfo {author} {\bibfnamefont {H.~C.}\ \bibnamefont {Freyhardt}}, \bibinfo {author} {\bibfnamefont {L.}~\bibnamefont {Fernández}}, \bibinfo {author} {\bibfnamefont {B.}~\bibnamefont {Holzapfel}},\ and\ \bibinfo {author} {\bibfnamefont {R.}~\bibnamefont {Feenstra}},\ }\bibfield  {title} {\bibinfo {title} {Simultaneous inductive determination of grain and intergrain critical current densities of {YBa$_2$Cu$_3$O$_{7-x}$} coated conductors},\ }\href {https://doi.org/10.1063/1.1639940} {\bibfield  {journal} {\bibinfo  {journal} {Appl. Phys. Lett.}\ }\textbf
  {\bibinfo {volume} {84}},\ \bibinfo {pages} {230} (\bibinfo {year} {2004})}\BibitemShut {NoStop}%
\bibitem [{\citenamefont {Palau}\ \emph {et~al.}(2007)\citenamefont {Palau}, \citenamefont {Puig}, \citenamefont {Obradors},\ and\ \citenamefont {Jooss}}]{palau2007simultaneous}%
  \BibitemOpen
  \bibfield  {author} {\bibinfo {author} {\bibfnamefont {A.}~\bibnamefont {Palau}}, \bibinfo {author} {\bibfnamefont {T.}~\bibnamefont {Puig}}, \bibinfo {author} {\bibfnamefont {X.}~\bibnamefont {Obradors}},\ and\ \bibinfo {author} {\bibfnamefont {C.}~\bibnamefont {Jooss}},\ }\bibfield  {title} {\bibinfo {title} {Simultaneous determination of grain and grain-boundary critical currents in {YBa$_2$Cu$_3$O$_7$}-coated conductors by magnetic measurements},\ }\href {https://doi.org/10.1103/PhysRevB.75.054517} {\bibfield  {journal} {\bibinfo  {journal} {Phys. Rev. B}\ }\textbf {\bibinfo {volume} {75}},\ \bibinfo {pages} {054517} (\bibinfo {year} {2007})}\BibitemShut {NoStop}%
\bibitem [{\citenamefont {Guth}\ \emph {et~al.}(2004)\citenamefont {Guth}, \citenamefont {Born},\ and\ \citenamefont {Jooss}}]{guth2004inhomogeneous}%
  \BibitemOpen
  \bibfield  {author} {\bibinfo {author} {\bibfnamefont {K.}~\bibnamefont {Guth}}, \bibinfo {author} {\bibfnamefont {V.}~\bibnamefont {Born}},\ and\ \bibinfo {author} {\bibfnamefont {C.}~\bibnamefont {Jooss}},\ }\bibfield  {title} {\bibinfo {title} {Inhomogeneous current distribution in wide high-temperature superconducting small-angle grain boundaries},\ }\href {https://doi.org/10.1140/epjb/e2004-00384-5} {\bibfield  {journal} {\bibinfo  {journal} {Eur. Phys. J. B}\ }\textbf {\bibinfo {volume} {42}},\ \bibinfo {pages} {239} (\bibinfo {year} {2004})}\BibitemShut {NoStop}%
\bibitem [{\citenamefont {Born}\ \emph {et~al.}(2004)\citenamefont {Born}, \citenamefont {Guth}, \citenamefont {Freyhardt},\ and\ \citenamefont {Jooss}}]{born2004self}%
  \BibitemOpen
  \bibfield  {author} {\bibinfo {author} {\bibfnamefont {V.}~\bibnamefont {Born}}, \bibinfo {author} {\bibfnamefont {K.}~\bibnamefont {Guth}}, \bibinfo {author} {\bibfnamefont {H.~C.}\ \bibnamefont {Freyhardt}},\ and\ \bibinfo {author} {\bibfnamefont {C.}~\bibnamefont {Jooss}},\ }\bibfield  {title} {\bibinfo {title} {Self-enhanced flux penetration into small angle grain boundaries in {YBCO} thin films},\ }\href {https://doi.org/10.1088/0953-2048/17/3/015} {\bibfield  {journal} {\bibinfo  {journal} {Supercond. Sci. Technol.}\ }\textbf {\bibinfo {volume} {17}},\ \bibinfo {pages} {380} (\bibinfo {year} {2004})}\BibitemShut {NoStop}%
\bibitem [{\citenamefont {Jing}\ \emph {et~al.}(2016)\citenamefont {Jing}, \citenamefont {Yong},\ and\ \citenamefont {Zhou}}]{jing2016influences}%
  \BibitemOpen
  \bibfield  {author} {\bibinfo {author} {\bibfnamefont {Z.}~\bibnamefont {Jing}}, \bibinfo {author} {\bibfnamefont {H.}~\bibnamefont {Yong}},\ and\ \bibinfo {author} {\bibfnamefont {Y.}~\bibnamefont {Zhou}},\ }\bibfield  {title} {\bibinfo {title} {Influences of non-uniformities and anisotropies on the flux avalanche behaviors of type-{II} superconducting films},\ }\href {https://doi.org/10.1088/0953-2048/29/10/105001} {\bibfield  {journal} {\bibinfo  {journal} {Supercond. Sci. Technol.}\ }\textbf {\bibinfo {volume} {29}},\ \bibinfo {pages} {105001} (\bibinfo {year} {2016})}\BibitemShut {NoStop}%
\bibitem [{\citenamefont {Jing}\ \emph {et~al.}(2017)\citenamefont {Jing}, \citenamefont {Yong},\ and\ \citenamefont {Zhou}}]{jing2017numerical}%
  \BibitemOpen
  \bibfield  {author} {\bibinfo {author} {\bibfnamefont {Z.}~\bibnamefont {Jing}}, \bibinfo {author} {\bibfnamefont {H.}~\bibnamefont {Yong}},\ and\ \bibinfo {author} {\bibfnamefont {Y.}~\bibnamefont {Zhou}},\ }\bibfield  {title} {\bibinfo {title} {Numerical simulation on the flux avalanche behaviors of microstructured superconducting thin films},\ }\href {https://doi.org/10.1063/1.4974000} {\bibfield  {journal} {\bibinfo  {journal} {J. Appl. Phys.}\ }\textbf {\bibinfo {volume} {121}},\ \bibinfo {pages} {023902} (\bibinfo {year} {2017})}\BibitemShut {NoStop}%
\bibitem [{\citenamefont {Kim}\ \emph {et~al.}(1962)\citenamefont {Kim}, \citenamefont {Hempstead},\ and\ \citenamefont {Strnad}}]{kim1962critical}%
  \BibitemOpen
  \bibfield  {author} {\bibinfo {author} {\bibfnamefont {Y.~B.}\ \bibnamefont {Kim}}, \bibinfo {author} {\bibfnamefont {C.~F.}\ \bibnamefont {Hempstead}},\ and\ \bibinfo {author} {\bibfnamefont {A.~R.}\ \bibnamefont {Strnad}},\ }\bibfield  {title} {\bibinfo {title} {Critical persistent currents in hard superconductors},\ }\href {https://doi.org/10.1103/PhysRevLett.9.306} {\bibfield  {journal} {\bibinfo  {journal} {Phys. Rev. Lett.}\ }\textbf {\bibinfo {volume} {9}},\ \bibinfo {pages} {306} (\bibinfo {year} {1962})}\BibitemShut {NoStop}%
\bibitem [{\citenamefont {Ziegler}\ and\ \citenamefont {Biersack}(1985)}]{SRIM}%
  \BibitemOpen
  \bibfield  {author} {\bibinfo {author} {\bibfnamefont {J.~F.}\ \bibnamefont {Ziegler}}\ and\ \bibinfo {author} {\bibfnamefont {J.~P.}\ \bibnamefont {Biersack}},\ }\bibinfo {title} {The stopping and range of ions in matter},\ in\ \href {https://doi.org/10.1007/978-1-4615-8103-1_3} {\emph {\bibinfo {booktitle} {Treatise on Heavy-Ion Science: Volume 6: Astrophysics, Chemistry, and Condensed Matter}}},\ \bibinfo {editor} {edited by\ \bibinfo {editor} {\bibfnamefont {D.~A.}\ \bibnamefont {Bromley}}}\ (\bibinfo  {publisher} {Springer US},\ \bibinfo {address} {Boston, MA},\ \bibinfo {year} {1985})\ pp.\ \bibinfo {pages} {93--129}\BibitemShut {NoStop}%
\end{thebibliography}%

\end{document}